
\input phyzzx
\tolerance=5000
\overfullrule=0pt
\twelvepoint
\nopubblock
\line{\hfill IASSNS-HEP-91/64}
\line{\hfill CALT-68-1764}
\line{\hfill HUTP-92/A003}
\line{\hfill December 1991}
\titlepage
\title{Quantum Hair on Black Holes}
\author{Sidney Coleman\foot{Research supported in part by
NSF grant
PHY-87-14654}}
\vskip.2cm
\centerline{\it Lyman Laboratory of Physics}
\centerline{\it Harvard University}
\centerline{\it Cambridge, MA. 02138}
\author{John Preskill\foot{Research supported in part by DOE
grant
DE-AC03-81-ER40050}}
\vskip.2cm
\centerline{\it Lauritsen Laboratory of High Energy Physics}
\centerline{\it California Institute of Technology}
\centerline{\it Pasadena, CA. 91125}
\author{Frank Wilczek\foot{Research supported in part by DOE
grant
DE-FG02-90ER40542}}
\vskip.2cm
\centerline{{\it School of Natural Sciences}}
\centerline{{\it Institute for Advanced Study}}
\centerline{{\it Olden Lane}}
\centerline{{\it Princeton, N.J. 08540}}
\endpage

\abstract{A black hole may carry quantum numbers that are {\it not} associated
with massless gauge fields, contrary to the spirit of the ``no-hair'' theorems.
We describe in detail two different types of black hole hair that decay
exponentially at long range.
The first type is associated with discrete gauge charge and the screening is
due to the Higgs mechanism.  The second type is associated with color magnetic
charge, and the screening is due to color confinement.  In both cases, we
perform semi-classical calculations of the effect of the hair on local
observables outside the horizon, and on black hole thermodynamics.  These
effects are generated by virtual cosmic strings, or virtual electric flux
tubes, that sweep around the event horizon.  The effects of discrete gauge
charge are non-perturbative in $\hbar$, but the effects of color magnetic
charge  become $\hbar$-independent in a suitable limit.  We present an
alternative treatment of discrete gauge charge using dual variables, and
examine the possibility of black hole hair associated with discrete {\it
global} symmetry.  We draw the distinction between {\it primary} hair, which
endows a black hole with new quantum numbers, and {\it secondary} hair, which
does not, and we point out some varieties of secondary hair that occur in the
standard model of particle physics.}

\endpage

\chapter{Introduction}

\section{Challenges for the quantum theory of black holes}
It has been claimed that existing
results on the quantum mechanics of black holes
require a modification of the fundamental laws of quantum
mechanics.\Ref\swhprD{S. W. Hawking, Phys. Rev. {\bf D14}
(1976) 2460.}
As we shall soon explain, we do not think this claim is
well founded.  Nevertheless, it is undeniable that
the behavior of black holes in quantum mechanics presents
conceptual challenges and opportunities that have not been
adequately
met.
Before we describe the concrete new results in this paper,
it seems
appropriate to describe our view of the bigger picture to
which they
belong.

In fact
the quantum mechanics of black holes presents two sets of
problems,
with different characters.  They might be called the
microscopic and
the macroscopic problems.
The microscopic problems concern the structure of very small
black
holes, with mass of order the Planck mass.
It is difficult, for several reasons, to imagine that
the description of such
holes (if they exist) could fail to
require a fully developed theory of quantum gravity.

The first and simplest reason is that as the
mass of a black hole approaches the Planck mass,
its Schwarzschild radius $2GM$ approaches its Compton radius
$\hbar/M$.
Thus irreducible quantum fluctuations in position, which
are of order
the Compton radius, render
the classical concept of the
horizon -- or indeed the classical concept of the
gravitational field
near the nominal location of the hole -- problematic.  In
this
regard it is
instructive to consider briefly the behavior of ordinary
elementary
point particles on
the other side of the dividing line, \ie\ with mass much
less than
the Planck mass.   The gravitational field of the point
particle
at the most naive level would be described by a
Schwarzschild solution,
with horizon at $R = 2GM = 2{\hbar M/ M_{\rm Pl.}^2}$.
However when
this $R$ is comparable to or less than the
Compton radius $R_{\rm Compt.} = {\hbar / M}$
the naive description is
quite inappropriate -- and we quite properly do not regard
such elementary
particles as black holes.  Rather, to calculate the
influence of the
gravitational interaction between such a particle and
another particle, we
simply calculate Feynman graphs for graviton exchange.
(Thought)-experimental attempts to
``see'' the nominal gravitational field at distance $R$ ,
by considering scattering at very small impact parameter
(and
therefore large
momentum and energy, of order $\hbar/R$),
are doomed to failure.  In such scattering,
the amplitude for production of many pairs becomes
large, and this situation simply cannot
be described in terms of scattering from
an external field.
This behavior is of course drastically different from what
one would
have for a classical
black hole, which simply absorbs particles incident at
impact parameter less than the Schwarzschild radius.  It is
not
at all clear
how to interpolate between them.

Another difficulty is that
as the mass of the hole decreases toward the Planck mass,
the loop expansion parameter for the gravitational
corrections
to the effective Lagrangian approaches unity near the
horizon,
even if we imagine the ultraviolet divergences of this
expansion
are cut off.
(See the discussion immediately below.)  Thus the
problem of understanding the non-perturbative behavior of
quantum gravity, which includes its ultraviolet behavior as
a sub-problem,
cannot be avoided.
At that point one must
either throw up one's hands or
(what in the current state of the art amounts to the same
thing)
appeal to string theory.

Neither of these difficulties arise for
black holes whose
mass is much greater than the Planck mass.  For such holes,
the Schwarzschild radius is much larger than the Compton
radius,
so that quantum uncertainty in position does not seriously
interfere
with the determination of the space-time geometry.
Also,
the curvature is small (relative to the inverse Planck
length)
near, and therefore of course external to, the horizon.
In this case,
a semiclassical treatment of gravity is quite plausible.
To be more precise, if we assume that higher-order
corrections
to the effective action for gravity have an effective
ultraviolet cut-off of order the Planck mass, then these
corrections will be small
near and external to the horizon.  For example
the contribution to the action from
a potential correction
term of the form
$R_{\alpha\beta\gamma\delta}R^{\alpha\beta\gamma\delta}$
to the Lagrangian density,
where $R^{\alpha\beta\gamma\delta}$ is the Riemann curvature
tensor,
will on dimensional grounds
occur with a coefficient $M_{\rm Pl.}^{-2}$.  Therefore,
since
the only relevant scale at the horizon is set by the
Schwarzschild
radius,
the contribution of this term to the equations of motion
will
be
of order
$\hbar^2 / (R_{\rm Schwarzschild}^2 M_{\rm Pl.}^2) =
(M_{\rm Pl.}/M)^2 << 1$
relative to the contribution from
the ordinary Einstein term near the horizon, and even
smaller
outside.
Since the region inside the horizon is causally disconnected
from
the exterior of the black hole, the occurrence of truly large
curvature
near the singularity inside the horizon is not directly
relevant to
the physics seen by external observers, and the higher order
corrections to the effective Lagrangian may be neglected
everywhere in the
physically relevant region.
(Of course, this discussion, since it appeals to
the classical description of the metric, is valid
only
semi-classically -- regarding the geometry
as approximately fixed -- and to all
orders
in $\hbar$, but not necessarily beyond.)

In the macroscopic regime of large black holes, interesting
effects can still arise
{}from {\it cumulative} effect of small
curvature over large volumes.
In this regime, even in the absence of a
workable complete theory of quantum gravity, one may
hope to do semiclassical and perturbative
calculations that have a high degree of
plausibility and yet present
interesting global features.

Indeed, striking phenomena have been found in this regime,
notably the Hawking evaporation\Ref\swhcmp{S. W. Hawking,
Comm. Math. Phys. {\bf 43} (1975) 199.}
and the Bekenstein-Hawking entropy\REF\beken{J. D.
Bekenstein, Phys. Rev. {\bf D7} (1973)
2333.}\refmark{\swhcmp,\beken}
of black holes.
The radiation, because of its approximately thermal nature,
suggests that a stochastic element enters
essentially into the description of
macroscopic black holes.
The most radical suggestion is that black hole evaporation
allows in principle the evolution of pure into mixed states,
which of course would violate the normal laws of quantum
mechanics.
For it seems clear that a black hole could be formed from
the
collapse of diffuse matter initially in a pure
quantum-mechanical state,
and if the subsequent radiation from the evaporating  black
hole
were accurately thermal the evolved
state would be mixed at later times.\refmark{\swhprD}
Strictly speaking it is
not true that the radiation is accurately thermal, if
for no other reason then because
the mass, and therefore
the nominal temperature, of the black hole changes with
time.
This is the simplest and most basic of all back-reaction
effects, which
correlate earlier with later radiation, but not the only
one.
Further correlations could in principle be calculated
order-by-order in $\hbar$, but it is most unlikely that they
are
adequate to avoid
the core of the problem --  the apparent threat to normal
quantum mechanics -- posed by the stochastic radiation.  The
core of the
problem is
qualitative: for the causal structure characteristic of a
classical black
hole, the region at spatial infinity outside the black hole
is not
a complete (backwards) Cauchy surface; it does not allow one
to uniquely
connect the past and the future (See Figure 1 and its
caption).

\FIG\cause{The causal structure of a classical
Schwarzschild black hole formed by collapse is clearly
displayed in its Penrose diagram, (a).  The distant future
outside the
black hole, represented by the solid boundary, is not
complete, in the
sense that knowledge of the wave function on this surface
does not
allow one to construct it throughout space-time.  It could
be completed
by including the horizon, or the singularity (dotted lines).
Thus one could have a unique, unitary connection between the
wave function given in the
distant past before the collapse and the wave function on
either completion.
However if we lose contact with our friendly observers on
the horizon,
or at the singularity, information is lost and the evolution
appears not
to be unitary.  The problem is especially acute if we
imagine that
the black hole, after formation, eventually evaporates
completely -- for
then these observers must eventually cease to exist, as
shown in (b).}

To pose the core problem in the clearest and most dramatic
way, suppose
the hole evaporates completely.
Then the information about the wave function
that flowed through the horizon seems to have
disappeared permanently, and thus
it appears that a pure state has evolved into
a mixed one.

There are several possibilities to avoid this
affront to quantum physics, which is perhaps
the most challenging conceptual problem posed by macroscopic
black hole quantum mechanics.  We shall now
briefly discuss four of them.

One possibility is that the radiation is
sufficiently correlated to be a pure state all by itself,
despite the fact that it appears almost totally uncorrelated
(that is,
the result of a slow thermal leak) in the standard
semi-classical
calculation.  Now because the evaporation of large black
holes is slow,
the putative
influence of earlier on later radiation must be reflected in
some
quasi-static property of the hole.  That is, the hypothesis
that
the radiation all by itself contains a enough information
to determine the state -- that none
is truly lost through the horizon -- requires that the black
hole should be capable of storing some accurate,
stable record of how it was formed and what it has radiated.
Until recently the
conventional wisdom has been that
this could not be true: that black holes have no
hair, that is no (or very few) internal states.  Indeed, it
has
been demonstrated
fairly rigorously that at the classical level there are very
limited
possibilities for hair,\Ref\wald{See, for example, R. Wald,
{\it General Relativity} (University of Chicago Press,
Chicago, 1984).} and that each possibility
requires the existence of suitable massless gauge
fields.\REFS\massvecbek{J. D. Bekenstein, Phys. Rev. {\bf D5}
(1972) 1239, 2403}\REF\teitel{C. Teitelboim, Phys. Rev. {\bf D5} (1972)
2941.}\refsend
However, the major point of this paper will be
to demonstrate that there are additional possibilities for
hair, when
the quantum nature of the black hole is taken into account.
We
certainly do {\it not} claim to have found enough hair to
solve the
main conceptual problem under discussion, but it is entirely
possible
that further analysis along these lines will uncover more.

A variant
on this possibility arises
if the black hole does
not evaporate completely.  Then it may leave
a stable remnant whose internal
state could be correlated with the state of the emitted
radiation.
This again requires
that the stable remnant should be capable of supporting lots
of hair.
In this regard it is appropriate to recall that the classic
stable
black holes (extreme Kerr-Newmann holes) are calculated to
have a very large entropy, proportional to
the area of the event horizon.\Ref\swhextreme{S. W. Hawking,
Phys. Rev. {\bf D13} (1976) 191.}  These calculations do not
identify the
quantum states which the entropy presumably is averaging
over.
These internal states represent another form of hair.
Because the entropy is calculated to be proportional to
the area of the event horizon, it is
tempting to speculate that the internal states
are associated with the state of the
horizon, regarded as a quantum-mechanical
object.\Ref\Hooft{G. 't Hooft, Nucl. Phys. {\bf B335} (1990) 138.}
However the exact nature
of this hair, and its relation to the quantum hair we shall
discuss in the
bulk of this paper, is unclear to us at present.

A third possibility is that in some real sense there is no
physical singularity.  Physical behavior at arbitrarily
large
space-time curvature, such as formally appears near the
black hole singularities, very plausibly brings in new
degrees of
freedom in addition to the ones familiar from our
low-energy,
low-curvature experience, which may drastically affect the
nature of the singularity.  This, in turn, can significantly
change the nature of the conceptual problems in the
quantum theory of black holes.  Suppose, for example, that
the
singularity becomes timelike and naked -- a
possibility realized
for a recently discovered class of
black holes (extreme dilaton black holes with
$a>1$).\REFS\GibMae{G. W. Gibbons, and K. Maeda, Nucl. Phys.
{\bf B298} (1988) 741.}\REF\Gar{D. Garfinkle, G. T.
Horowitz, and A. Strominger, Phys. Rev. {\bf D43} (1991) 3140.}\refsend
In the dilaton black holes there
is a timelike singularity visible from infinity, and to
define the
quantum theory of fields in such a geometry
appropriate boundary conditions must be imposed at the
singularity.
These boundary conditions are not uniquely fixed by the
macroscopic theory,
but would be determined by the underlying microscopic theory
(\eg\ superstring theory) needed to describe regions
of truly large curvature.
If the boundary conditions are correctly chosen, it is
entirely
reasonable to suppose that unitarity may be maintained.
There is
a well-known precedent for this situation,  in the behavior
of
gauge theory magnetic monopoles (Callan-Rubakov
effect).\Ref\Rub{V. Rubakov, JETP Lett. {\bf 33} (1981) 644;
C. Callan, Phys. Rev. {\bf D25} (1982) 2141.} In that
case, at the level of the effective field theory of
electromagnetism,
the monopole has a singularity at the origin.  Spin-$1\over
2$
fermions with the minimal charge consistent with the
Dirac quantization condition, {\it viz.\/} ${eg} = {1\over
2}$,
feel no centrifugal barrier and in s-wave scattering reach
the
center with finite probability.  Suitable boundary
conditions can be
supplied, so that the S-matrix describing this problem is
unitary.
One can also describe non-singular monopoles in a
non-abelian extension
of the low-energy model.  Then the scattering problem is
entirely
well defined.  Nevertheless, the description of low energy
scattering
in the extended theory can be accurately described by the
effective
low energy theory; the rich additional physics of the full
theory, in
this limited context merely serves to fix the boundary
conditions.
For {\it space-like} singularities such
as those which occur for conventional black holes, however,
boundary conditions would amount to {\it constraints} on the
form of
the wave  function, which are presumably not physically
sensible.
(How does the initial wave function know it is going to
describe
collapse to a black hole, and
had better acquire intricate non-local correlations?)

The only apparent problem with this
possibility is that it requires rather special field
content,
and even so occurs only for extreme charged black holes.
Thus it seems unlikely to be relevant to describe
really large uncharged black holes, for which
a self-consistent description
in terms of low-mass fields outside the horizon
ought to be a good approximation, unless there are low-mass
fields
(\eg\ axions, dilatons)
whose existence has for some reason eluded observation  to
date.

Finally,
perhaps the most straightforward way out is to deny that
information
is truly lost down the singularity.
As a model which is close to classical, and
thus easy to discuss, suppose for example that the field
content of the theory is somehow altered so that high
curvature
induces a rebound of the metric, so that the deep interior
of the
black hole rather than containing a singularity opens up
into
another nonsingular space -- this will
be the spore of a ``baby universe.''
(Note that the singularity theorems, which say that
singularities are inevitable, always assume constraints on
the energy
momentum tensor which are not true for generic forms of
matter, and
certainly need not apply to the effective Lagrangian at
extreme
curvatures.)
This
spore is necessarily causally disconnected from our own, at
the classical
level, because it is behind the horizon.
Information about the wave function flowing through the
horizon is now stored in the spore, rather than being lost
at a singularity.
It is then
conceivable that the final act in the complete evaporation
of the
black hole involves the spore pinching off, to form a
separate, self-contained
baby universe.\Ref\baby{S. W. Hawking, Phys. Rev. {\bf D37}
(1988) 904;
S. W. Hawking and R. Laflamme, Phys. Lett. {\bf 209B} (1988)
39.}
The baby universe may or may not resemble a full-scale
universe
with an interesting cosmology.  Also the pinching off
process may or may not be rare,
depending on unknown details of the matter content -- it is
even conceivable
that the typical black hole gives birth to many babies.

Do these possibilities help us avoid
the evolution of pure into mixed states?  At first
sight it seems they give no help at
all, but rather sharpen the problem.
We have thrown information into the baby universe, where it
is forever lost
to us.  In other words there are correlations between the
state of the
baby universe and the state of the parent universe, and
therefore either
alone is described by a mixed state.
However, we must recognize that the strict separation of
baby and parent
is an approximation valid
only at the classical level. {\it If a baby universe can
branch off
quantum mechanically, it can also come back quantum
mechanically}.
This simple but crucial remark implies that the proper
description of the
final state must include the state of the baby together with
the parent;
even if we inhabit the latter, we are not allowed to ignore
the former.
The penalty for this form of child neglect
is revocation of license to practice
quantum mechanics.

This discussion is closely related to the
arguments made by one of us
concerning the description of wormhole processes as
effective interactions.\REFS\coleman{S. Coleman, Nucl. Phys.
{\bf B307} (1988) 867; {\bf B310} (1988) 643.}\REF\giddings{S.
Giddings and A. Strominger, Nucl.
Phys. {\bf B307} (1988) 854.}\REF\swhworm{S. W. Hawking, Mod. Phys. Lett
{\bf A5} (1990) 145, 453.}\refsend  While the
wormhole process might be roughly considered as the birth of
a baby universe,
in a more careful discussion one must consider it as
coherent emission into a ``bath''
of baby universes, a process whose macroscopic
phenomenological consequences in our
universe can be captured in an effective vertex,
whose precise form depends on the wave function describing
the bath.  In this description, the difference between what the black
hole absorbs and what it emits is correlated with the baby universe wave
function, and quantum coherence is explicitly maintained.

Thus, to summarize, it does not seem that one is forced to
conclude
that the process of black hole evaporation violates or
transcends
the normal laws of quantum mechanics.  There are several
ways whereby
one can
imagine reconciling the stochastic element of black hole
evaporation, and the apparently irreversible flow of
information through
the horizon, with the principles of quantum mechanics (and a
unitary
time evolution.)  Some, though not all, of these
possibilities require
the existence of new, essentially quantum-mechanical,
internal states -- quantum hair -- for black holes.

\section{Black holes and elementary particle properties}

Besides the fundamental challenge to the principles of
quantum mechanics
mentioned above, the known behavior of black holes creates a
certain tension
in the description of matter.  In the standard
semiclassical treatment of a black hole, the hole is
described
as a thermal object with very few internal degrees of
freedom.
The normal description of an elementary particle is very
different -- no temperature appears,
and the particle may have many internal quantum numbers.
Yet one might like to believe that there is no fundamental
distinction
between these forms of matter; that a sufficiently heavy
elementary particle
(any one heavier than the Planck mass) would in fact be a
black hole.

In order to reconcile these descriptions, one necessary step
is
certainly to show that black holes are capable of carrying
internal quantum
numbers, contrary to the spirit of the classic no-hair
theorems.  That is
what we shall accomplish below.

One more or less plausible possibility for black holes in
the real
world is precisely that there actually are
elementary particles of this sort.
For example, if electromagnetic gauge symmetry is first
unified into
a compact gauge group at an extremely large
(super-Planckian) mass scale,
then the stable magnetic monopoles will be black holes.
These could
be formed in the big bang, either directly or as the
remnants or
radiation products of the evaporation of other mini-black
holes.
Our considerations show that there are additional
possibilities -- that
tiny black holes can be stabilized against Hawking
evaporation
by other gauge charges, associated
with {\it broken} gauge groups.  Thus
if we are ever lucky enough to encounter
a stable mini-black hole, the
reason for its stability might not be at all obvious.  And
indeed there
could well be a rich spectrum of such objects, with
different masses,
magnetic charges, and discrete charges.

\section{Summary of the content}
The content of the rest of this paper is as follows.  In
Section 2
we discuss the reasoning leading to the classic (and
classical)
no-hair theorems of black hole physics.  We then argue on
physical
grounds that these theorems must be violated in the case of
discrete
gauge charge.  Concrete models embodying this physics, and
the
related physics of discrete magnetic charge, are exhibited.
In Section
3 we discuss in careful detail the quantum treatment of the
ordinary
Reissner-Nordstr\"om black hole.  The formal implementation
of the charge projection is fully discussed, and shown to be
accomplished by
a weighted integral over field configurations with specified
``vorticity'' (line integral $\oint d\tau A_\tau$, where
$\tau$ is
the periodic imaginary time variable).  The
formal difference between electric
and magnetic charge is emphasized.   In Section 4 this
apparatus is
adapted to the case of discrete electric gauge charge.  It
is shown that
the effect of discrete gauge charge is embodied in an
imaginary time
vortex configuration, and is non-perturbative in $\hbar$.
It is argued that this configuration corresponds to the
space-time
process (discussed in Section 2)
whereby a virtual cosmic string wraps around the black hole,
and is
in this sense to be interpreted as a world-sheet instanton
for the
cosmic string.  Several
charge-dependent physical observables, including
non-vanishing electric fields outside the horizon and
corrections
to the classic relation between temperature and mass for the
hole, are
calculated in appropriate limits.  In Section 5 we discuss
the
corresponding physics for discrete magnetic charge.  It is
argued that
although there are drastic formal differences between
electric and
magnetic charge, it is quite plausible that their physical
behavior is not
drastically different.
In Section 6  we discuss the dual description of broken
symmetry phases.
This material
has nothing to do with black holes {\it per se}, but
the material is not entirely
standard and is used in the following section.
One simple but striking result is that in the dual
description of the
Higgs phase, where the phase of the scalar field is
represented in terms
of
a rank-two antisymmetric tensor field $B$,
the charge coupling takes the
form of an interaction $B\wedge F$ that resembles a
$\theta$-term.
In Section 7 we discuss
an alternative approach to discrete gauge hair more along
the lines
presented in the literature as ``axion charge'', and attempt
to clarify
the relationship between these approaches.
Finally in Section 8 we draw the distinction between primary
and secondary
quantum hair.  We give examples of secondary hair within the
standard model
of particle physics, and emphasize that the discrete hair
discussed in the
bulk of the paper is {\it primary}.

We shall adopt units in which $c=1$, but will display
factors of $G$ and $\hbar$.

\def\pd#1#2{{{\partial #1}\over {\partial #2}}}
\chapter{Discrete Gauge Hair: Preliminary Discussion}
In this chapter we shall review the conceptual framework
for no-hair theorems, and then discuss why these theorems
are not expected to apply in the case of discrete gauge
hair.
Several of the topics discussed here in physical terms will
be discussed more formally in the following chapters.

\section{No-hair theorems}

To understand the essence of the no-hair
theorems,\Ref\price{R. H. Price, Phys. Rev. {\bf D5}
(1972) 2419, 2439.} it is
best to
begin by considering the simplest case of
a scalar field in the background of a Schwarzschild black
hole.
The metric, in Schwarzschild coordinates, is
$$
ds^2 = \left(1-{2GM\over r}\right)dt^2 -
\left(1-{2GM\over r}\right)^{-1}dr^2 - r^2d^2\Omega
\eqn\discaa
$$
where, of course, $d^2\Omega$ is the line element on the
unit sphere.
In analyzing the wave equation in this background, it is
convenient
to introduce the ``tortoise coordinate''
$$
r_* \equiv r + 2GM \ln [(r - 2GM)/2GM]
\eqn\discab
$$
which satisfies
$$
dr_* = \left(1-{2GM\over r}\right)^{-1} dr \, .
\eqn\discac
$$
Because the metric at fixed angle is proportional, in this
variable, to that of flat space:
$$
ds^2_{~{\rm fixed angle}} = \left(1-{2GM\over r}\right)
(dt^2 - dr_*^2) \, ,
\eqn\discad
$$
the wave equation will assume a particularly simple form.

It is crucial to note that $r_*$ has the properties
$$
\eqalign{
r_*  \rightarrow r + 2GM\ln r \, ; ~~~~~  &r \rightarrow
\infty \cr
r_*  \rightarrow \ln (r - 2GM) \rightarrow - \infty \,
;~~~~~
&r \rightarrow 2GM\, .\cr}
\eqn\discae
$$
The first of these equations
implies that $r_*$ reduces essentially to $r$, and
in particular that it becomes positively infinite, as
$r\rightarrow \infty$.
(The extra logarithm here is due to the long-range nature of
the
gravitational interaction, and directly reflects the
``Coulomb logarithm''
in the phase of waves at infinity.)  More important for our
present
considerations is the second equation.  It tells us that
$r_*$, the
natural variable for the wave equation, approaches negative
infinity
at the horizon.  In this sense, the horizon is ``infinitely
far away.''

The wave equation is easiest to analyze by inserting the
partial wave
expansion for the scalar field $\Psi$
$$
\Psi (t,r,\Omega ) = {1\over r} \sum_{l,m} \psi_{l,m}
(t,r_*) Y_{lm}(\Omega )
\eqn\discaf
$$
into the wave equation.
One obtains
$$
\left(-\pd{^2}{t^2} + \pd{^2}{r_*^2} \right) \psi_{l,m} =
\left(1 - {2GM\over r} \right)
\left(\mu^2 + {2GM\over r^3} + {l(l+1)\over r^2}\right)
\psi_{l,m}
\eqn\discag
$$
where $\mu$ is the mass (really the inverse Compton
wavelength) of the scalar.
The right-hand side of this equation may be interpreted as
an effective
potential.  This effective potential vanishes near the
horizon
(\ie\ as $r_* \rightarrow -\infty$), approaches simply
$\mu^2$ as
$r_* \rightarrow \infty$, and is everywhere positive.

Now consider the zero-frequency limit.
Clearly, because the
second derivative never changes sign, a solution that
decreases exponentially at infinity will have to blow up
at $r_* \rightarrow -\infty$, \ie\ at the horizon.  Such
behavior
is physically unacceptable: it involves infinite energy in
the
$\Psi$ field, among other things.  Thus, there are no
physically
acceptable static solutions; the only acceptable static
scalar field
configuration is identically zero.\refmark{\massvecbek}
In particular there can be no exponentially falling
Yukawa tails, as would occur far from
a normal point source coupled to $\Psi$.  A black hole
cannot be a source: it has no hair.

The crucial circumstance underlying this result is clearly
the fact
that the natural variable
$r_*$ is unbounded in both directions, approaching negative
infinity
at the horizon.  It is for this reason that continuation of
any possible
tail from spatial infinity costs not merely a large, but
inevitably
an infinite, amount of energy.

Price\refmark{\price} analyzed wave equations for higher
integer spin fields as well.   Unfortunately these equations are
most conveniently analyzed using the Newman-Penrose\Ref\newpen{E.~Newman,
R.~Penrose, J. Math. Phys. {\bf 3} (1962) 566.} formalism,
which may unfamiliar to many readers.  Furthermore the
equations, although satisfactory in flat space, are inconsistent in
curved space for spin $s\geq 3$.
Thus there is more than one large hole appearing
in the analysis.  However since the question of higher-spin hair is
potentially very important, we will now briefly summarize and
make some tentative
observations on the analysis.

Using the Newman-Penrose formalism,
and extending the flat-space equations formally
by minimal coupling, one arrives at equations of the form
$$
{d^2\over dr_*^2}(r^{s+1}\hat\Phi_0) - F_l(r_*)r^{s+1}\hat\Phi_0 = 0
\eqn\zza
$$
for the static part of the $l^{\rm th}$ partial wave
of the fundamental quantity $\hat\Phi_0$.  Actually $\hat\Phi_0$
is the coefficient of a spherical harmonic $Y_{lm}$, and there are
$2l+1$ independent components, which we shall leave implicit.
(Eq.~\zza\ is derived by simple manipulation of two particular
components of the minimally coupled Newman-Penrose system of
equations.  If we had chosen other components and done more complicated
manipulations, we would have gotten a different equation for
$\hat\Phi_0$ -- this is the inconsistency mentioned above.  However
any ``reasonable'' manipulation leads to an equation of the same
general form as \zza\ , but with a modestly modified function
$F_l$.)
Here
$$
F_l \equiv (1-2GM/r){l(l+1)\over r^2}.
\eqn\zzb
$$

$\hat\Phi_0$ is the de-spun field of zero conformal and spin weight.
The two crucial properties of $\hat\Phi_0$ are that:

\pointbegin
The ordinary spinor or tensor components of the fields may be obtained
from $\hat\Phi_0$ by differentiation and algebraic manipulations.

\point
$\hat\Phi_0$ is directly physically meaningful, and
is expected to be
well-behaved both at the horizon and at infinity.
For example in the
electromagnetic case $s=1$ one finds
$\hat\Phi_0 = -{1\over 2}(E_r + iB_r)$.  Because the field strengths
(and not the potentials) occur here, $\hat\Phi_0$ is physically meaningful.
Furthermore because the radial components of the field strength are
unaffected by boosts in the radial direction, the physical requirement
that freely falling observers see no singular behavior at the horizon
entails that $\hat\Phi_0$ must be non-singular at the horizon.
Of course,  manifestly it must be well-behaved at infinity.

\par

Taking eq.~\zza\ at face value, one sees that there is a fundamental
distinction between $l\leq s-1$ partial waves and the higher ones.
In both cases one has the asymptotics $\hat\Phi_0 \rightarrow
a_l + b_lr_*$
for solutions near the horizon
$r_* \rightarrow -\infty$, with only $b_l=0$ being physically
acceptable.  As $r_* \rightarrow \infty$ the solutions behave as
$\hat\Phi_0 \rightarrow c_lr_*^{l-s} + d_lr_*^{-l-s-1}$, and here
there is a big difference.  For $l\ge s$, this solution is physically
acceptable only if $c_l=0$.
But one may repeat the argument given for the case
of scalar fields to show that the solution that is nonsingular at the event
horizon ($a_l\ne 0, ~b_l=0$) matches up with a solution with $c_l\ne 0$ at
$r_*=\infty$.  Therefore, the only allowed static solution has $\hat\Phi_0$
identically zero -- there is no hair.  However, for the lower partial waves,
this argument does not apply, and acceptable non-trivial solutions do indeed
exist.

The characteristic feature of
these allowed solutions is that the field carries
a non-vanishing value of a conserved surface integral.  Propagation of
radiation to spatial infinity
cannot change the value of this surface integral.
Since no causal process, including in particular the formation of a black
hole, can flout this conservation law, one expects that the conserved surface
integral corresponds to a variety of
hair that can reside on a stationary black
hole.

As we have mentioned, the full set of equations for higher spin fields in
curved space is inconsistent.  However, as Price argues, it is quite
plausible that the (unknown) corrections that lead to
consistent equations, while they will alter the form of the
effective potential and lead to mixing among modes of the same
quantum numbers, will not alter the crucial asymptotic behaviors
near the horizon and spatial infinity.
This is because at spatial infinity one should have something
arbitrarily close to the flat space equations, while
close to the horizon the effective potential is always killed by
the $1-2GM/r$ stretching factor.

When mass terms are introduced into the field equations, the analysis of
classical hair on stationary
black holes is qualitatively altered.  This can be
discussed most cogently in the spin-1 case, where no issue of consistency
arises.  For a massive spin-1 field, there is no conserved surface integral,
and hence no compelling reason for spin-1 hair (that is, electric or magnetic
charge) to reside on a black hole.
Indeed, the analysis of the field equations
shows that the physically acceptable solution at the horizon fails to match up
with a decaying exponentially at
infinity, so that only the trivial solution is
allowed.\refmark{\massvecbek,\teitel}  It is instructive to consider what
happens, in the case of a vector field with mass $\mu$, in the limit $\mu\to
0$.  When a charged particle that acts as a source for a massive vector field
falls through the horizon of a black hole,
the vector field outside the horizon
leaks away in a (Schwarzschild) time of order $\mu^{-1}$.  Thus, as $\mu$ gets
smaller, the time scale for the decay of the field gets longer and longer;
finally, in the limit $\mu\to 0$, the field persists indefinitely.

Apart from the question of consistency, at the level of formal
manipulation and asymptotic behavior the
Newman-Penrose-Price equations for higher spins do not look
very different in principle from those for spin 1.
In the body of this paper we shall analyze spin 1 in depth, and
show that soft mass terms (arising from spontaneous breakdown of
the gauge invariance associated with massless spin-1 fields) do not
destroy the hair entirely, if there is a
remnant discrete gauge symmetry left behind.  In many ways the
crucial ingredient of the argument is the construction of appropriate
charge projections; and this part could be carried through for
higher spin too.  However the physical interpretation of the hair
only emerges clearly when one can implement a Higgs mechanism and
construct vortices; this might be doable for higher spins at least in
flat space, and presents an attractive subject for
further investigation.

A simple but perhaps not entirely misguided way to regard
the
no-hair theorems is as follows: in collapse to a black hole,
any information that
{\it can\/} fall through the horizon {\it will\/} fall
through
the horizon.  The universal action of gravitation
sweeps clean; and
all trace of what has been swept in disappears, because once
beyond the
horizon it loses causal contact with the exterior.
The only quantities that escape destruction are those which
can be
seen from the safe distance of infinity; that is, those
connected with
conserved surface integrals, such as the ones we have
discussed above.

\section{Discrete gauge symmetry}

Upon first hearing, the notion of local discrete symmetry in
the continuum
may sound rather silly.  Indeed, the most important
dynamical consequence
of a continuous
local symmetry is the existence of a new field, the gauge
field.  This
field arises
when one introduces a gauge potential, in order to
formulate covariant derivatives.  Covariant derivatives are
of course
necessary, so that invariant interactions involving
gradients may be
formed.  Such interactions in turn are necessary in order
that charged
fields may propagate.
In the case of a discrete symmetry there is no similar need
to
introduce a gauge potential,
because the ordinary derivative already transforms simply.

Alternatively, since any path can be continuously
deformed to a trivial path, the corresponding
parallel transport can be continuously deformed to the
identity.
But for a discrete group this means we must have the
identity all along:
thus all parallel transport is trivial.
Also, if the discrete gauge group arises as the remnant
of an initially continuous group, after
all continuous symmetries have been
wiped out by the Higgs mechanism, one is accustomed to think
that charge is completely screened.

Upon further reflection, however, one realizes that
each of these arguments has serious limitations.
First, there are situations in which one cannot deform all
paths
to the identity.  This occurs, by definition, on spaces that
are not
simply connected.
Second, there are situations in which the parallel
transport is not a continuous function of the path, or is
ill-defined for
some paths.  An important class of examples, which will be
extensively
considered below, concerns flux tubes in the Higgs phase of
gauge theories.
In that case the gauge symmetry ``expands'' to the full
unbroken
symmetry at the core of the flux tube.
Parallel transport through this core will
not in general stay within the discrete subgroup, so from
the
point of view of the low-energy theory (which does not
take account of the expanded gauge symmetry) there is a
singularity.

Likewise, the statement that in a Higgs phase
charge is completely screened
is too broad.  One must appreciate
that the screening occurring in the Higgs mechanism is not
a mystical
process, but essentially a special case of the homely
phenomenon of
dielectric polarization
as occurs in ordinary dielectrics or plasmas.  It is true
that
we are concerned,
in the Higgs mechanism, with a particularly effective
screening, produced by
a condensate of massless charged particles.   Such particles
are capable of
producing a dipole distribution, to exactly cancel any
imposed electric field.
The particles in the condensate transform trivially under
the remnant
discrete symmetry, however, and it is unreasonable to expect
that any
distribution of them can perfectly mimic a non-trivial
source.

There is a profound difference between local and global
symmetries, whether continuous or discrete.  While
global symmetry is a statement that the laws of physics take
the same
form when expressed in terms of various distinct variables,
local
symmetry is a statement that the variables used in a
physical theory
are redundant.
In language that may be more familiar, this redundancy is
often
stated as the fact that in a gauge theory, only gauge
invariant
quantities are physically meaningful.
{}From this point of view, it is clear that no processes,
not even such exotic
ones as collapse to a black hole or black hole evaporation,
can violate a
gauge symmetry.
On these very general grounds, then, we should expect that
a black hole must be  capable of carrying discrete gauge
charges, reflecting the charges of what fell in to make it.
However, these abstract arguments leave it totally
unclear what the concrete physical implications of discrete
gauge
charges could be.

\section{Discrete electric gauge hair}

To address this question, and to make the whole discussion
more concrete,
let us now consider a specific realization of the general
idea of discrete
local symmetry.\Ref\krawil{L. M. Krauss and F. Wilczek,
Phys. Rev. Lett. {\bf 62} (1989) 1221.} Consider a U(1)
gauge theory containing two
complex scalar fields
$\eta$, $\xi$ carrying charge $Ne$, $e$ respectively.
Thus we have for the Lagrangian
$$
{\cal L} =  -{1\over 4}F_{\mu\nu}F^{\mu\nu}
+ |(\partial_\mu  + iNeA_\mu ) \eta |^2
+ |(\partial_\mu + ieA_\mu ) \xi )|^2 - V(|\eta |~,|\xi |)
\, .
\eqn\discba
$$
Suppose that $\eta$
undergoes condensation at some very high mass scale $v$,
while $\xi$
produces quanta of relatively small mass and does not
condense.

Before the condensation the
theory is invariant under the local gauge
transformations
$$
\eqalign{
\eta (x)'  &= \exp (iNe\Lambda (x)) \eta (x) \cr
\xi (x)'  &= \exp (ie\Lambda (x))  \xi (x) \cr
A_\mu (x)' &= A_\mu (x) - \partial_\mu \Lambda (x) \, ,\cr }
\eqn\discbb
$$
where $\Lambda$ is a real variable.
However the condensate characterized by the vacuum
expectation value
$$
<\eta (x)> = v
\eqn\discbc
$$
in the homogeneous ground state is invariant only when
$\Lambda$ is an integer multiple of $2\pi/Ne$.  These
residual
transformations still act non-trivially on $\xi$,
multiplying
it by various $N^{\rm th}$ roots of unity.  Their possible
actions
generate the discrete group $Z_N$.

The
effective low energy theory well below
$v$ will simply be the theory of the
single complex scalar
field $\xi$; neither the gauge field nor
$\eta$ will appear in the effective theory,
since these fields produce very massive quanta.
The only implication of the original gauge symmetry for the
low energy
effective theory is then the absence of interaction
terms forbidden by the residual discrete symmetry.
(If there were additional charged scalar fields in the
theory, the
discrete symmetry would forbid many couplings that were
otherwise possible.)
But this
implication,
so far, does not distinguish between local and global
symmetry.

Once we widen our horizons to consider processes occurring at
energies
of order $v$, of course the underlying gauge degrees of
freedom,
if they exist, can be
excited.  A more subtle manifestation is also possible.
The broken symmetry theory
contains stable strings threaded by magnetic flux
2$\pi$/$Ne$,
as follows.
The theory supports vortex solutions, where the $\eta$ field
behaves asymptotically as a function of the angle $\theta$
as
$$
{\eta (r,\theta) \rightarrow  ve^{-i\theta},\  r \rightarrow
\infty }~,
\eqn\discbd
$$
where $v$ is the magnitude of $\eta$ in the homogeneous
ground state.
Along with this asymptotic behavior for $\eta$ we must have
for
the gauge potential
$$
{A_{\theta}(r, \theta) \rightarrow {1\over {Ne}} }
\eqn\discbe
$$
so that the covariant derivative $D_{\theta}\eta =
(\partial_{\theta} + iNeA_{\theta} )\eta$,
which appears (squared) in the energy density,
vanishes at infinity.

In this set-up the field strength also vanishes
asymptotically.  Indeed we note that
by making the choice $\Lambda = \theta/Ne$, we can
formally remove the space
dependence of $\eta$ in \discbd\ and make $A_{\theta}$ in
\discbe\
vanish altogether.  In so doing, we have
(formally) transformed back to the homogeneous
ground state.
However of course the
gauge transformation function
$\Lambda$ is not quite kosher, since the angle
$\theta$ is not a legitimate single-valued function.  The
correct statement
is that the asymptotic behavior of the vortex
is trivial and can be gauged away locally,
but not globally.  Since we can pick a well-defined branch
of
$\theta$ in any
patch that does not surround the origin, all local gauge
invariant
quantities must reduce to their ground state values -- this
explains, if you
like, why $D\eta$ and $F$ vanish.  But the line integral of
$A$ around
a closed loop surrounding the origin, which according to
Stokes measures
the flux inside, cannot be changed by any legitimate gauge
transformation,
and it is definitely {\it not} zero for the vortex.  Indeed
we find the
basic flux unit is $\Phi_N =  {2\pi / Ne} $.

Another perspective on the
global non-triviality of the vortex, is that our putative
gauge
transformation $\Lambda = \theta/Ne$ transforms a unit
charge field
such as
$\xi$ into something that is not single-valued: following
\discba\ we
find that $\xi' (\theta + 2\pi) = \exp ({2\pi i\over N})
\xi' (\theta)$.

What has all this got to do with discrete gauge theories in
the continuum?
Well, the condensate $<\eta > = v$ is not invariant under a
general gauge
transformation, but it is invariant under the discrete
subgroup
generated by $\Lambda = {2\pi/ Ne}$.  This discrete subgroup
acts
trivially both on $\eta$ and, it would seem,
on the gauge field ($\Lambda$, since it takes
only discrete values, cannot change continuously at all).
However,
$\Lambda$ and its various powers are definitely not trivial
acting on
$\xi$, which gets multiplied by powers of the $N^{\rm th}$
root of unity.
Thus
there is a discrete but non-trivial gauge subgroup left.
Moreover,
{\it the gauge transformation associated with winding around
a vortex is
precisely an element of the residual discrete gauge symmetry
group} -- this
is just a restatement, in our new interpretation, of the
result of the
previous
paragraph.  Alternatively we could say that the Wilson loop
for parallel
transport around the vortex defines an element of the
residual gauge group:
$$
{\exp (\oint ie A_{\mu} dx^{\mu} ) \ \epsilon \ Z_N .}
\eqn\discbf
$$
We anticipated that the discrete gauge group might come into
its own in the
presence of singularities.  Now we see how this is quite
simply realized.
The required ``singularity'' is here supplied by the vortex
core.  In that
core, where the condensate vanishes,
the discrete gauge symmetry blows up into a full scale
continuous
$U(1)$, and the vector potential is unfrozen.  The only
trace of the vortex
visible outside the vortex core is the total
flux -- which is none other than an element
of the residual discrete gauge symmetry group.

The scattering
of $\xi$ quanta, with charge
$e$, from such strings\Ref\Rohm{R. Rohm, Princeton
University Ph.D. thesis (1985) unpublished; M. G. Alford and
F. Wilczek, Phys. Rev. Lett. {\bf 62} (1989) 1071.} is
dominated, at low energies,
by the Aharonov-Bohm
effect.\Ref\AharBo{Y. Aharonov and D. Bohm, Phys. Rev. {\bf
119} (1959) 485.}  The magnitude and form
of this cross-section is uniquely
determined by the product of charge and flux, modulo
2$\pi\hbar$, and thus
allows one, in principle, to make a precise
observational determination of the $Z_{N}$-valued charge
alluded
to above.  Of course postulating the existence of such
strings takes
us outside the framework of the effective low-energy theory
as usually
understood, so the existence of this effect does not really
contradict
the statement made above, that the low-energy
effective theory does not distinguish
local from global discrete symmetries.

\REF\bowick{M. J. Bowick, S. B. Giddings, J. A. Harvey,
G. T. Horowitz, and A. Strominger, Phys. Rev. Lett. {\bf 61}
(1988) 2823.}
A simple thought experiment\refmark{\krawil}
based on the Aharonov-Bohm
scattering
process provides a heuristic but
convincing demonstration that black holes have discrete
gauge hair.\foot{The concept of black hole hair that is
detected via Aharonov-Bohm interactions was first discussed
in Ref. \bowick.}
For let us imagine that we have a $\xi$ quantum falling into
a black
hole, and scatter a string of very low energy and momentum
{}from this
composite object.  The scattering cross-section, which
involves behaviors
at large times and distances, should not depend on the
precise instant
at which the particle crosses the event horizon - a rather
fuzzy notion,
in any case.  And yet this cross-section does depend
critically on the $Z_{N}$ charge.  We must conclude that
this $Z_{N}$
charge does not depend on whether the particle has crossed
the event
horizon, and in particular that it retains its meaning (and
induces the
same Aharonov-Bohm phases) for the asymptotic, ``pure''
black hole.
A more formal proof, whose core idea is really the same,
could be based on the discrete analog of Gauss'
law, which insures that the $Z_{N}$ charge inside a volume
can be
expressed in terms of the expectation values of quantum
field operators
on the bounding surface.\REFS\AMW{M. G. Alford,
J. March-Russell,
and F. Wilczek, Nucl. Phys. {\bf B337} (1990)
695.}\REF\prekra{J. Preskill and L. M. Krauss, Nucl. Phys.
{\bf B341} (1990) 50.}\refsend

We would now like to discuss why this discrete
electric gauge hair is properly
{\it quantum}  hair, with no
ordinary classical limit.  In a sense this is obvious from
the
nature of screening.  In a Higgs theory, there is a
condensate
of massless charged particles in the ground state, and at
the classical level
this condensate
is ready and willing to screen any test charge.  The only
thing that
prevents perfect screening is the quantization of the charge
of the
condensate particles.  Thus, the survival of any
consequences of discrete electric
gauge hair must depend on quantization.  To
better appreciate this, it
is helpful to understand more explicitly how the charge
units vary with
$\hbar$.  The classical action, obtained by integrating
the Lagrangian density \discba\ over space and time, must
have the
units of $\hbar$.  From this it follows easily that the $e$
appearing
in the Lagrangian has units of (action)$^{-{1\over 2}}$.
For clarity
we shall write this quantity as $e_f$, ($f$ for field
charge).
In the ordinary classical limit, $e_f$ is held fixed. On the
other
hand the charge density operator of the theory, which
multiplies $A_0$ in the
Lagrangian, is of the generic form
$i e_f \phi^\dagger {\buildrel \leftrightarrow
\over \partial_0} \phi$.
Because the fundamental commutation
relation normalizes a bilinear form in $\phi$ and
$\partial_0 \phi$ to
$\hbar$, it follows that the charge operator is essentially
$\hbar e_f$
times the number operator.  Thus the charge $e_p$
($p$ for particle) of a single quantum is
$$
e_p = \hbar e_f \, .
\eqn\discbg
$$
We see that electric charges which are only finite multiples
of
$e_p$ vanish in the ordinary classical limit.  (It will be
useful to remember that $\hbar e_f^2$
and  $e_p^2 / \hbar$ are dimensionless quantities.)

It would be both disturbing and disappointing if the only
manifestation
of discrete quantum hair on a black hole involved explicit
cosmic strings.
Among other things, this would close off some of the ideas
mentioned in the
introductory chapter, which require hair capable of
influencing ordinary
particles.  And indeed if one has a real process sensitive
to the charge, it
seems quite reasonable to expect that there are associated
virtual processes
also sensitive to the charge.  Thus we might expect that the
discrete charge manifests itself as an instruction
concerning
how to weight the
relative phase of amplitudes for the black hole wrapping one
way
or the other around an appropriate cosmic string (or, of
course, the
string wrapping around the hole).  Upon changing real to
virtual,
we expect that {\it the
discrete electric charge on the hole should instruct us how
to weight
the phase of amplitudes for processes
where virtual cosmic string loops  nucleate, envelop the
hole,
and re-annihilate.}  Described in other words,
the relevant processes are ones in which virtual
string world sheets wrap around
the hole.\REFS\preskill{J. Preskill, Phys. Scrip. {\bf T36}
(1991) 258.}\REF\cpwA{S. Coleman, J. Preskill, and F.
Wilczek, Mod. Phys. Lett. {\bf A6} (1991) 1631.}\REF\cpwB{S.
Coleman, J. Preskill, and F. Wilczek, Phys. Rev. Lett. {\bf
67} (1991) 1975.}\refsend

In the following chapters, we shall verify this expectation
in great detail.

The virtual string world sheet
process is of course a tunneling process,
and will be non-perturbative
in $\hbar$; indeed it will be exponentially small in
$1/\hbar$.
(Note that cosmic strings are classical objects: their
structure follows from solving the classical field
equations, and so their flux and
size remain fixed as $\hbar \rightarrow 0$ with $e_f$ and
$v$ fixed.)
This contrasts with what one might expect for an ordinary
particle
carrying discrete charge, where there
are small charge-dependent effects (say in the scattering of
two
charged particles) of order $e_p^2$ from the
short-range Yukawa fields.
The effect for black holes is different from, and much more
subtle than, restoration of their lost Yukawa tails.
Another important
difference is that for black holes, but not for ordinary
particles,
discrete hair expands the space of states -- see chapter
8.

\section{Discrete magnetic hair}

Now let us discuss the magnetic analog of discrete electric
gauge
hair.  At the most naive level one might expect that these
are
trivially related, because of the well-known duality
property of
the free Maxwell equations.  And indeed in the context of
classical
black hole physics, the difference between electrically
and magnetically charged holes
is entirely trivial -- both are described by essentially the
same Reissner-Nordstr\"om, or Kerr-Newman, solutions.
However
the formal treatment of these
objects in the semiclassical limit must be quite different,
as
we shall see in the next chapter.  One circumstance that
makes
this plausible is that the electric charge quantum is
$e_p = \hbar e_f$ while the magnetic charge quantum is
simply
$e_f^{-1}$, so the latter but not the former is finite in
the classical
limit.  Nevertheless, at the end of the day the properties,
specifically
the radius and temperature,  of
an electric hole whose charge is many quanta are closely
similar
to those of the magnetic hole with the same
magnitude of field strength and mass.

What about broken symmetry phases?
If a gauge theory is in a weak coupling Higgs phase,
electric fields are
screened, and magnetic fields are confined to flux tubes.
These phenomena can be described classically---the screening
length and flux quantum are independent of $\hbar$.  If a
gauge theory is in a confinement phase, magnetic fields are
screened, and electric fields are confined to flux tubes.
These phenomena are quantum--mechanical; in fact, the
inverse screening length is non-perturbative in $\hbar$.

We have seen that $Z_N$ electric charges can be introduced
into a Higgs theory, such that the charges have an
Aharonov--Bohm interaction with a magnetic flux tube.
The construction of weakly coupled
models with $Z_N$ magnetic monopoles
that can be studied with semiclassical methods is a bit more
involved.  (There is an enormous literature concerned with
discrete
gauge theory monopoles at strong coupling, which are often
invoked
as an explanation of confinement; but as far as we know
these
ideas have never been pursued very far within the framework
of a
well controlled approximation.)

Probably the simplest model, at least conceptually,
is based on the spontaneous breakdown of the
gauge group $SO(N^2 -1)$ to $SU(N)/Z_N$.
This may be accomplished by a Higgs field in the
three-index antisymmetric tensor representation,
$H_{ijk}$, that acquires a vacuum expectation value
$$
<H_{ijk}> = vf_{ijk}
\eqn\disca
$$
proportional to the structure constants $f_{ijk}$ of
$SU(N)$.
Note that $SU(N)$ is of dimension $N^2 -1$.  It may be
regarded
as a subgroup of $SO(N^2-1)$, because it can be represented
as
the group of orthogonal transformations (leaving invariant
the trace of Casimir operator, or the Killing form) acting
by conjugation on its Lie algebra.
The condition that such a transformation is an isomorphism
of the
Lie algebra of $SU(N)$ is precisely that it leaves the
structure constants
invariant.  The group of automorphisms of the Lie algebra of
$SU(N)$ is $SU(N)/Z_N$, acting by conjugation.  (The central
elements,
of course, do not generate non-trivial automorphisms this
way.)
Thus we see that $SU(N)/Z_N$ is nicely embedded in $SO(N^2
-1)$, and
that there is a simple way of breaking the larger group down
to the
smaller one.\Ref\london{E. J. Weinberg, D. London, and J.
Rosner, Nucl. Phys. {\bf B236} (1984) 90.}

In this framework we may identify configurations
carrying $Z_N$ magnetic flux, as follows.
The homogeneous ground state defined by \disca\ is not
unique;
one may obtain states that are equally good energetically by
acting
on it with any $SO(N^2-1)$ transformation.  Of course in
this
procedure the $SU(N)/Z_N$ is powerless, and the manifold of
formally unequal (but gauge equivalent) candidate ground
states is the coset space $SO(N^2-1)/(SU(N)/Z_N)$.  Now
let us suppose, as usual in constructing gauge theory
monopoles,\Ref\polyakov{A. M. Polyakov, JETP Lett. {\bf 20}
(1974) 194; G. 't Hooft, Nucl. Phys. {\bf B79} (1974) 276.}
that the symmetry may break to different points within this
manifold, depending on the direction in space, at spatial
infinity.  Now consider the parallel transport around a
sequence of
loops, each starting and ending at the north pole; the loop
begins as an infinitesimal loop, expands to one which is a
great circle through
the south pole, and then comes back to another small loop,
having
lassoed the sphere once.
Since each of these loops
starts and ends at the north pole, the total parallel
transport
defined by each must leave the Higgs field at
the north pole invariant; thus it must define an element of
$SU(N)/Z_N$.
Furthermore the small loops at the beginning and end must
give the
identity element in $SU(N)/Z_N$.  Thus the sequence of
parallel transports,
as the loops lasso the sphere,
defines an element of the homotopy group
$\pi_1 (SU(N)/Z_N) = Z_N$.  The different homotopy classes
cannot
be connected to one another by continuous changes over a
finite volume.
Thus
there is infinite energy barrier between them, and,
minimizing the
energy within each sector, one will find some
stable monopole configurations.
(Of course the doubly charged pole may in principle decay
into two
singly-charged poles, and so forth.)

Alternatively, the flux may be defined by patching
hemispheres.  On a hemisphere, one may perform an $SO(N^2-
1)$ gauge
transformation to  rotate the vacuum expectation values so
that
they are all pointing in the same direction.
Likewise, one may align the other hemisphere.  However
matching
the gauge transformations along the intersection of these
hemispheres
will introduce a closed path in $SU(N)/Z_N$, as before.

The $Z_N$ magnetic charge, then, can be defined by
essentially
{\it classical} operations at infinity.  In the black hole
sector too,
one should be able to find classical solutions
with the specified asymptotic flux.  (And later, we will.)

Thus we see that there is a magnetic analogue of discrete
electric hair,
though superficially it appears to differ from it
qualitatively.
However, on closer consideration one finds that the
difference
is not entirely sharp.  Suppose, for example (not
unrealistically),
that the $SU(N)/Z_N$ theory we have been discussing so far
classically, actually confines electric charge
(and screens magnetic charge) at large distances.
Then strictly speaking
there will not be any magnetic flux at infinity.  It might
seem, then,
that the magnetic black hole has been shaved bald.  However
in the
confined phase one has {\it electric flux tubes}, analogous
to the magnetic
flux tubes of the Higgs phase, because now it is electric
flux
that is confined.
These flux tubes will be able to detect the screened
discrete magnetic
charge on the black hole, by the dual of the Aharonov-Bohm
process,
scattering off electric flux tubes.\REF\erice{S. Coleman,
in {\it The Unity of the Fundamental Interactions}, ed. A.
Zichichi (Plenum, New York, 1983).}\REF\susskind{M. Srednicki
and L.
Susskind, Nucl. Phys. {\bf B179} (1981)
239.}\refmark{\prekra,\coleman,\susskind}  This physics, of
course,
is entirely
analogous to what we have seen in the Higgs phase.

(To avoid misunderstanding, we should probably emphasize
that such screened magnetic hair does {\it not} occur in the
standard model, with unbroken gauge group $[SU(3)_{\rm
color}\times U(1)_{\rm em}]/Z_3$.  Though this model admits
magnetic monopoles that carry a $Z_3$ color magnetic flux,
there are no stable electric flux tubes -- the tube can
break via nucleation of a quark-antiquark pair.  Hence there
is no means of detecting $Z_3$ magnetic charge at long
range.  Furthermore, and not coincidentally, the $Z_3$ color magnetic charge
of a monopole is completely determined by its $U(1)_{\rm em}$ magnetic
charge;  it is not an independent quantum number.
If black holes can carry screened magnetic hair
in Nature, this hair is not associated with the known strong
interaction.  Rather, it must be associated with another, as
yet unknown, confining gauge interaction that admits genuine
$Z_N$ monopoles.)

The real difference between
the two cases is not qualitative, but
only quantitative.   Confinement is non-perturbative
in $\hbar$, so that classically
a magnetic charge generates magnetic flux at infinity;
in a straightforward weak-coupling analysis,
there
is visible hair.
Subtleties arise only after one realizes that there is an
important
effect at large distances, confinement,
which is non-perturbative
in $\hbar$.  For if one then attempts to visualize this
effect classically,
or to incorporate it into an improved effective Lagrangian
which one then
treats classically, the classic (classical) no-hair argument
will
come into play.  However, fortunately, the more
straightforward
electric screening case has alerted us that the hair
remains,
because of
another nonperturbative effect in $\hbar$, namely the
wrapping of real or virtual flux tubes around the hole.
Thus if confinement is regarded as a strong effect,
rather than as a tiny correction, the magnetic
charge case comes to look very similar to the electric
charge case.

\chapter {The Reissner--Nordstr\"om Black Hole}

\section{The Euclidean Path Integral}
One of our objectives is to extend the standard
semiclassical analysis of black hole thermodynamics to
include effects that are nonperturbative in $\hbar$.
Nonperturbative effects are most conveniently studied using
Euclidean path integral methods, and we will use such
methods here.  In fact, we do not know another way to obtain
our main results.

We employ the Euclidean path integral with reservations,
because in the context of black hole physics, the
foundations underlying this formalism are not completely
secure.  There are both technical and conceptual problems.
The main technical problem is that the  Euclidean
Einstein--Hilbert action is unbounded from below.  The
integral over
the conformal degree of freedom of the metric must be
defined with some care, and it is not clear what the correct
prescription is.  The main conceptual problem arises because
the Euclidean formalism, as we use it, applies to a black
hole in thermal equilibrium with a surrounding radiation
bath, rather than a black hole evaporating into empty space.
The equilibrium is typically unstable if the radiation bath
is infinite in extent, because the black hole has negative
heat capacity.  This difficulty can be avoided by enclosing
the radiation in a sufficiently small reflecting cavity.
The deeper question is whether the concept of a black hole
in equilibrium with radiation makes sense.  A stationary
spacetime filled with radiation does not satisfy the
Einstein equations.  The back reaction of the radiation on
the geometry  causes the spacetime to evolve, so that it
becomes unclear how the concept of thermal equilibrium can
apply.  (There is, of course, an approximate notion of
thermal equilibrium in the semiclassical limit, since the
temperature of the radiation is of order $\hbar$.)

We will ignore these problems in this paper.  Our working
hypothesis is that the notion of black hole thermodynamics,
and the use of the Euclidean path integral to probe the
thermal behavior, are sensible, at least in a semiclassical
approximation.

In this section, we will illustrate the Euclidean method by
applying it to the case of a black hole that carries
electric or magnetic charge (the Reissner--Nordstr\"om black
hole).  We particularly want to emphasize how charge
projection operators are inserted into the partition
function, for the purpose of studying the thermodynamics of
a particular charge sector.  Similar charge projections will
be invoked in the subsequent two sections, which treat the
cases of {\it screened} electric and magnetic charge.

\section{The Semiclassical Limit}
We are interested in the thermodynamic behavior of a black
hole in the semiclassical limit, the limit of small $\hbar$.
To define this limit precisely, we must specify what
quantities are to be held fixed as $\hbar$ approaches zero.

In a model of electromagnetism coupled to gravity, there are
two coupling constants, Newton's constant $G$ and the
electromagnetic gauge coupling $e$.  Both of these are to be
regarded as classical quantities; that is, they are held
fixed as $\hbar\rightarrow 0$.  We emphasize (again) that
$e$ denotes here the coupling that appears in the classical
action, and has the dimensions of $(action)^{-1/2}$.  Thus,
$e$ is related to the electric charge $e_{\rm particle}$ of
an elementary particle by
$$
e_{\rm particle}=\hbar e~.
\eqn\semiA
$$

The semiclassical limit of a black hole with mass $M$ and
electric charge $Q$ is defined by holding both $M$ and $Q$
fixed as $\hbar\rightarrow 0$.  Thus, in this limit, the
mass becomes arbitrarily large compared to the Planck mass
$(\hbar /G)^{1/2}$, and the charge becomes arbitrarily large
compared to the charge quantum $\hbar e$.  In terms of black
hole thermodynamics, the length scale $\beta\hbar$ is held
fixed, where $\beta^{-1}$ is the black hole temperature.  In
the semiclassical limit, then, this length scale (the
typical wavelength of a thermal radiation quantum) becomes
arbitrarily large compared to the Planck length $(\hbar
G)^{1/2}$.

\section{The Electric Charge Projection}

We will now describe how the Euclidean path integral method
is used to evaluate the partition function in a particular
charge sector of an abelian gauge theory.  At first, we will
consider quantum field theory on flat spacetime (no
gravity).  The extension to include gravity will be
discussed later.

Let us briefly recall the standard method of deriving the
path integral
expression for the partition function of a gauge
theory.\Ref\gross{D. J. Gross, R. D. Pisarski, and L. G. Yaffe, Rev.
Mod. Phys. {\bf 53} (1981) 43.}  We
work in the
the gauge $A_0 = 0$ and evaluate
$$
Z(\beta) = {\rm tr}\left( e^{-\beta H}\right)
\eqn\projA
$$
by summing over the basis of eigenstates of $A_i(\vec x)$
and $\phi(\vec x)$ (where $\phi$ denotes the matter fields).
However,
since only physical states are to be included in the sum, we
must also
insert a projection onto states that satisfy the Gauss law
constraint.
We thus obtain
        $$Z(\beta) = \int ~dA_i(\vec x) ~d\phi(\vec x)
{}~d\Omega(\vec x) \VEV{A^\Omega,\phi^\Omega |
        e^{-\beta H}|A,\phi}~,
        \eqn\projB
        $$
where $\Omega(\vec x)$ is a (time--independent) local gauge
transformation.  It is important to notice that, since the
Gauss law
constraint only requires that physical states are invariant
under gauge
transformations of compact support, $\Omega(\vec x)$ is
restricted in
eq.~\projB\ to obey
        $$\Omega(\vec x) \to {\bf 1} {\rm ~~as~~} |\vec x|
\equiv r \to
        \infty~.
        \eqn\projC$$

Now up to a factor of the volume of the local gauge group
(which must be
removed by gauge fixing), eq.~\projB\ may be reexpressed as
        $$Z(\beta) = \int_{\beta\hbar} dA_\mu~d\phi
\exp\left(-S_E[A,\phi]/\hbar\right)~,
        \eqn\projD$$
where the histories $A_\mu(\tau,\vec x),\phi(\tau,\vec x)$
are required to be periodic in
Euclidean time $\tau$ with period $\beta\hbar$.
Furthermore, as a
consequence of eq.~\projC, $A_\tau$ must satisfy
        $$A_\tau (\tau,\vec x) \to 0 {\rm ~~as~~} r \to
\infty~.
        \eqn\projE$$
(We can recover eq.~\projB\ and eq.~\projC\ from eq.~\projD\
and eq.~\projE\ by imposing the temporal gauge condition
$A_\tau=0$.)

The partition function $Z$ defined by eq.~\projD\ includes a
sum over {\it
all} physical states.  But if the Hilbert space of the
theory contains
superselection sectors, we may wish to consider $Z$
restricted to  a single
sector.  This requires that further projection operators be
inserted in
the path integral.

Let us specialize now to the case of an abelian gauge theory
(in the
Coulomb phase), and construct the partition function
restricted to the
states of specified electric charge $Q$.  A state in the
charge--$Q$
sector satisfies
        $$U(\Omega)\ket{Q} = e^{i\omega Q/\hbar e}\ket{Q}~,
        \eqn\projF
        $$
where the gauge transformation $\Omega(\vec x)$ has the
asymptotic form
        $$\Omega(\vec x) \to e^{i\omega} = {\rm constant,~~
as} ~ r \to \infty~,
        \eqn\projG$$
and $U(\Omega)$ is the unitary operator that represents this
gauge transformation.
The projection operator onto physical states of charge $Q$
is
        $$P_Q = \int^{2\pi}_0 ~ {{d\omega}\over 2\pi} ~
\int~
        d\Lambda(\vec x) ~ e^{-i\omega Q/\hbar e}
{}~U\left(e^{i(\Lambda + \omega)}
        \right)~,
        \eqn\projH$$
where
        $$e^{i\Lambda(\vec x)} \to {\bf 1}~~{\rm as}~~ r \to
\infty  ~.
        \eqn\projI$$
Thus, the partition function in the charge--$Q$ sector may
be expressed
as
        $$\eqalign{
        Z(\beta,Q) &= {\rm tr} \left(P_Q e^{-\beta
H}\right)\cr
        &= \int^{2\pi}_0 ~ {{d\omega}\over 2\pi} ~ e^{-
i\omega Q/\hbar                 e}~
        \int~d\Lambda(\vec
x)\VEV{A-{1\over e}\partial\Lambda,
e^{i(\Lambda+\omega)}\phi|e^{-\beta
H}|     A,\phi}\cr
        &= \int_0^{2\pi} ~ {{d\omega}\over 2\pi} ~e^{-
i\omega Q/\hbar e} ~\hat Z(\beta,\omega)\cr}
{}~.
        \eqn\projJ$$
Here,
        $$\hat Z(\beta,\omega) = \int_{\beta\hbar,\omega}
dA_\mu~d\phi ~\exp\big(-S_E[A,\phi]/\hbar\big)
        \eqn\projK$$
is (up to gauge fixing) the Euclidean path integral over
configurations
that are periodic in $\tau$ with period $\beta\hbar$, and
also
satisfy the
constraint
        $$\exp\left(ie\int^{\beta\hbar}_0 ~d\tau
{}~A_\tau(\tau,\vec x)\right) = e^{i\omega}~,
        \quad{\rm for}\quad r = \infty~.
        \eqn\projL$$
Note that eq.~\projL\ constrains only the noninteger part of
$(e/2\pi)\int d\tau A_\tau$.  By combining the sum over the
integer part with the $\omega$ integral, we may extend the
range of integration to $(-\infty,\infty)$, and rewrite
eq.~\projJ\ as
$$
Z(\beta,Q)=\int_{-\infty}^\infty{d\omega \over 2\pi}
e^{-i\omega Q/\hbar e} Z(\beta,\omega)~,
\eqn\projM
$$
where now $Z(\beta,\omega)$ is defined so that the gauge
field is restricted by
$$e\int_0^{\beta\hbar} d \tau \, A_{\tau}(\tau,\vec x)~
 \Big|_{r = \infty} = \omega~.
 \eqn\projN$$
This is our result for the partition function in a sector of
specified charge.


\section{The Schwarzschild Black Hole}
Let us now briefly review the Euclidean path integral
analysis of the thermodynamics of an uncharged black hole in
the semiclassical limit, as originally performed by Gibbons
and Hawking.\Ref\gibhawk{G. W. Gibbons and S. W. Hawking,
Phys. Rev. {\bf D15} (1977) 2752.}

If thermal fluctuations in the  geometry are to be included,
eq.~\projD\ can be extended to include an integral over the
(Euclidean) spacetime metric.  The integral over metrics
divides into topologically distinct sectors.  The
``trivial'' sector, which includes small fluctuations about
flat space, is probed by integrating over Euclidean
geometries that have the topology $R^3\times S^1$.  Gibbons
and Hawking argued that the thermodynamics of a black hole
with temperature $\beta^{-1}$ can be studied by summing over
geometries that have topology $R^2\times S^2$, are
asymptotically flat, and are periodic in imaginary time
$\tau$ with period $\beta\hbar$.

To check this hypothesis, we should verify that the standard
results of black hole thermodynamics can be recovered from
this prescription in the semiclassical limit.  In this
limit, the integral
$$
Z(\beta)\equiv e^{-\beta F}=\int_{\beta\hbar}e^{-S_E/\hbar}
\eqn\schwarzA
$$
in a given topological sector is dominated by the solution
to the Euclidean field equations in that sector that has the
lowest Euclidean action.  Our task, then, is to find that
solution and to evaluate its action.

The crucial observation, now, is that the Euclidean section
of the Schwarzschild geometry has the $R^2\times S^2$
topology, and is periodic in imaginary time.  To see this,
consider the somewhat more general case of a static
spherically symmetric Euclidean geometry, with metric
$$
ds^2=e^{2\Phi}d\tau^2+e^{2\Lambda}dr^2 + r^2d\Omega^2~,
\eqn\schwarzB
$$
where $\Phi$ and $\Lambda$ are functions of $r$ only.
Suppose that $e^{2\Phi}>0$ and $e^{2\Lambda}>0$ for $r>r_+$,
and $e^{2\Phi}=0$ for $r=r_+$.  Thus, $r=r_+$ is the
location of an event horizon in the Lorentzian continuation
of this geometry.  In the vicinity of $r=r_+$ this metric
can be rewritten in the form
$$
ds^2=\left((e^{\Phi})'e^{-\Lambda}~\big|_{r=r_+}\right)^2
R^2 d\tau^2 + dR^2 + r^2d\Omega^2~,
\eqn\schwarzC
$$
where $R=0$ at $r=r_+$, and the prime denotes
differentiation with respect to $r$.
We see that $\tau$ can be interpreted as the angular
coordinate on the $R-\tau$ plane.  But a singularity at
$R=0$ can be avoided only if $\tau$ is a periodic variable
with period $\beta\hbar$, where
$$
2\pi(\beta\hbar)^{-1}=(e^{\Phi})'e^{-
\Lambda}~\big|_{r=r_+}~.
\eqn\schwarzD
$$
If this condition is satisfied, then the Euclidean metric
eq.~\schwarzB, with $r\geq r_+$ is topologically $R^2\times
S^2$; it is the analytic continuation of that part of the
Lorentzian geometry that lies outside or at the event
horizon, and the two--sphere at the origin $R=0$ is the
horizon two--sphere.  By computing the proper acceleration
of static observers near the event horizon of the Lorentzian
geometry,
and using eq.~\schwarzD,
we find
$$
\kappa\equiv a_{\rm proper}~e^\Phi=2\pi/\beta\hbar~,
\eqn\schwarzE
$$
which is the relation between the surface gravity $\kappa$
and the black hole temperature $\beta^{-1}$ discovered by
Hawking.\refmark{\swhcmp}

In the special case of the Euclidean Schwarzschild solution
we have
$$
e^{2\Phi}=e^{-2\Lambda}= 1-{2GM\over r}~;
\eqn\schwarzF
$$
thus, $r_+=2GM$, $\kappa=1/4GM$, and $\beta\hbar=8\pi GM$.

Now we turn to the evaluation of the action of this
solution.  This calculation is somewhat delicate, for two
reasons.  First, the Euclidean Einstein--Hilbert action
includes a boundary term, and the action of the
Schwarzschild solution arises solely from this boundary
contribution.  Second, the boundary term is formally
infinite; to define it, we must perform an appropriate
subtraction.

The gravitational action has the form
$$
S_{\rm grav} = - {1 \over 16 \pi G} \, \int \, d^4x \sqrt g
\, R +
S_{\rm boundary}~.
\eqn\schwarzG
$$
The first term evidently vanishes for a solution to the
vacuum Einstein equations, or for a solution such that the
stress tensor of the matter has a vanishing trace.  In such
cases only the boundary term contributes.

For later reference, we will describe the evaluation of the
boundary term for the general static, spherically symmetric
geometry, eq.~\schwarzB.  We will suppose that the geometry
is asymptotically flat, and will take the boundary to be
$r=\infty$.  But since a subtraction will be required, we
will first suppose that the boundary is at a finite radius
$r$, and will take the $r\to \infty$ limit only after
subtracting.

In this case, the boundary term can be written as
$$
S_{\rm boundary}=-{1\over 8\pi G}~\partial_{\rm
normal}~({\rm
volume~ of~ boundary})~;
\eqn\schwarzH
$$
this is the rate, per unit of proper distance, at which the
volume of the boundary increases as the boundary is
displaced in a direction along the outward--pointing vector
normal to the boundary.  The ``boundary'' at radius $r$ is
an $S^1 \times S^2$ with
$$
{\rm volume}=\left( e^{\Phi (r)}\beta\hbar\right)\left(4\pi
r^2\right)~,
\eqn\schwarzI
$$
and so we find
$$
S_{\rm boundary}=\left(-{\beta\hbar\over 2G}\right)e^{-
\Lambda (r)}
\left( r^2 e^{\Phi (r)}\right)'~.
\eqn\schwarzJ
$$
This expression diverges linearly as $r\to \infty$.

The prescription given by Gibbons and Hawking is to subtract
the action of a flat spacetime with the same induced
geometry at the boundary.  In the present case, the flat
spacetime has the topology $R^3\times S^1$, and the proper
circumference of the circle is $e^{\Phi (r)}\beta\hbar$.
Thus, the ``temperature'' of the flat spacetime matches the
red shifted
temperature that would be measured by a static observer at
radius $r$ in
the curved spacetime.

The matching flat geometry has action
$$
S_{\rm boundary}^{(flat)}=\left(-{\beta\hbar\over 2G}\right)
e^{\Phi (r)}(r^2)'~.
\eqn\schwarzK
$$
If the functions $e^{2\phi}$ and $e^{2\Lambda}$ appearing in
the metric have the asymptotic large-$r$ expansions
$$
e^{2\Lambda}\sim 1+{2A\over r}~,~~~e^{-2\Phi}\sim 1+{2B\over
r}~,
\eqn\schwarzL
$$
then the subtracted action is
$$
S_{\rm boundary}-S_{\rm boundary}^{(flat)}=-{\beta\hbar\over
2G}(B-2A)~.
\eqn\schwarzM
$$
Finally, if eq.~\schwarzB\ asymptotically solves the vacuum
Einstein equations, then $A=B=GM$, where $M$ is the ADM
mass, and we have
$$
S_{\rm boundary}-S_{\rm boundary}^{(flat)}={1\over
2}\beta\hbar M~.
\eqn\schwarzN
$$

Now we perform the semiclassical evaluation of the free
energy of a Schwarzschild black hole by saturating the path
integral with the Euclidean Schwarzschild solution.  We have
$$
e^{-\beta F}\simeq e^{-S_{\rm Schwarzschild}/\hbar}~,
\eqn\schwarzO
$$
or
$$
\beta F(\beta)={1\over 2}\beta M= 4\pi {(GM)^2\over \hbar G}
={\hbar\over 16\pi G }\beta^2~,
\eqn\schwarzP
$$
where we have used the relation $\beta\hbar=8\pi GM$ that
was derived from eq.~\schwarzD. From this equation, we may
also obtain the entropy
$$
{\rm entropy}=\beta M -\beta F(\beta)=4\pi(GM)^2/\hbar G~,
\eqn\schwarzQ
$$
which agrees with the Bekenstein--Hawking value.

\section{The Electrically Charged Black Hole}
We now proceed to the semiclassical evaluation of the
thermodynamic functions for an electrically charged black
hole.

The action is now
$$
S=S_{\rm grav}+S_{\rm em}~,
\eqn\elecA
$$
where
$$
S_{\rm em}= {1 \over 16 \pi} \, \int d^4 x\, \sqrt g \,
F_{\mu \nu}
\,F_{\lambda \sigma}\, g^{\mu \lambda} g^{\nu \sigma}~.
\eqn\elecB
$$
(We ignore, for now, the charged matter fields.)  The
stationary point of the action in the sector with $R^2\times
S^2$ topology that is periodic in imaginary time with period
$\beta\hbar$ and obeys the constraint eq.~\projN\ is the
Euclidean Reissner--Nordstr\"om solution, namely
$$
A_\tau={\omega\over \beta\hbar e}\left(1-{r_+\over
r}\right)~,~~
F_{r\tau}={\omega\over \beta\hbar e}{r_+\over r^2}~,
\eqn\elecC
$$
$$
\eqalign {
e^{\Phi+\Lambda} &= 1 \, , \cr
e^{\Phi -\Lambda} &= 1 - \Big[ 1 - G\left({\omega \over
\beta\hbar e}\right)^2\Big]
\left({ r_+
\over r}\right) -
G\left({\omega \over \beta\hbar e}\right)^2
\left({ r_+ \over r}\right)^2~.\cr}
\eqn\elecD
$$
{}From eq.~\schwarzD\ we determine $r_+$ to be
$$
r_+={\beta\hbar\over 4
\pi}\left[1+G\left({\omega\over\beta\hbar e}
\right)^2\right]~.
\eqn\elecE
$$
Since the electromagnetic stress--tensor is traceless, the
gravitational contribution to the action arises solely from
the boundary term; evaluating it using eq.~\schwarzM, we
find
$$
\eqalign{
S_{\rm grav}-S_{\rm grav}^{(flat)}&={\beta\hbar\over 4G}
\left[1-G\left({\omega\over\beta\hbar
e}\right)^2\right]r_+\cr
&={(\beta\hbar)^2\over16\pi G}\left[1-G^2\left({\omega\over
\beta\hbar e}\right)^4\right]~.\cr}
\eqn\elecF
$$
The electromagnetic contribution to the action is
$$
\eqalign{
S_{\rm em}&={1\over 2}\beta\hbar\left({\omega\over
\beta\hbar e}
\right)^2r_+\cr
&={(\beta\hbar)^2\over 8\pi}\left({\omega\over\beta\hbar e}
\right)^2\left[1+G\left({\omega\over\beta\hbar
e}\right)^2\right]
{}~.\cr}
\eqn\elecG
$$
Combining the two contributions gives
$$
S-S^{(flat)}={(\beta\hbar)^2\over 16\pi
G}\left[1+G\left({\omega
\over\beta\hbar e}\right)^2\right]^2~.
\eqn\elecH
$$

Now we project out the contribution from the charge--$Q$
sector by integrating over $\omega$ as in eq.~\projM,
obtaining
$$\exp[- \beta F(\beta, Q)] = \int_{-\infty}^\infty
{d\omega\over 2\pi} \exp
\left[- i {\omega Q\over\hbar e}\right]$$
$$ \times~~ \exp \left\{- {1\over\hbar} ~
{(\beta\hbar)^2\over 16 \pi G}
\left[1 + G \left({\omega\over\beta\hbar
e}\right)^2\right]^2\right\} ~. \eqn\elecI$$
In the semiclassical limit, $\hbar$ approaches zero with
$\beta\hbar$,
$Q$, $G$, and $e$ held fixed.  Thus, the $\omega$ integral
can be
evaluated in the steepest--descent approximation.  The
$\omega$-contour
is deformed so that it passes through a saddle point on the
imaginary
axis.  (The contour is {\it not} rotated; its ends must be
fixed, or
else $Z(\beta,\omega)$ will blow up.)  The semiclassical
expression for
the free energy becomes
$$
F(\beta, Q) = (\Omega(\beta, \Phi) + Q\Phi)\big|_{{\rm
stationary}} ~,
\eqn\elecJ
$$
where we have defined a new dummy variable
$$
\Phi = {i\omega\over\beta\hbar e} ~,
\eqn\elecK
$$
and
$$
        \Omega(\beta,\Phi) = {\beta\hbar\over 16\pi G} (1 -
G\Phi^2)^2 ~.
\eqn\elecL
$$
We see that eq.~\elecJ\ may be interpreted as a Legendre
transform;
$\Phi$ is the ``chemical potential'' coupled to the electric
charge $Q$,
and $\Omega(\beta,\Phi)$ is the associated thermodynamic
potential.

The value of $\Phi$ at the saddle point is $\Phi=Q/r_+$;
this can be
interpreted as the
electrostatic potential difference between the black hole
horizon and
spatial infinity, for a black hole with electric charge $Q$
and radius
$r_+$.  In fact,
the solution given by eq.~\elecC-\elecE, with $\Phi$
assuming this
saddle--point value, is the analytic continuation to
imaginary time of
the  Reissner--Nordstr\"om black hole solution with
electric charge $Q$.  This continued solution has imaginary
$F_{r\tau}$,
and negative electromagnetic action.

An alternative way to describe the semiclassical
calculation, then, is
as follows:  If the electrically charged Reissner--Nordstrom
black hole
solution is continued to imaginary time, the Euclidean
action of the
continued solution, divided by $\beta\hbar$, is the
thermodynamic
potential $\Omega(\beta,\Phi)$ of the black hole.  This is
the result
found by Gibbons and Hawking.\refmark{\gibhawk}

The more familiar formulas of black hole thermodynamics can
be recovered from eq.~\elecL\ by Legendre transformations
and changes of variable.  For example, we may write
$$
\beta\Omega={1\over\hbar}\left(S-S^{(flat)}\right)={1\over
2}\beta
( M-Q\Phi)~,
\eqn\elecM
$$
where the first term is the gravitational action and the
second term is the electromagnetic action. ($M$ is the black
hole mass.)  Therefore,
$$
\beta F= {1\over 2}\beta (M + Q\Phi)=
{1\over\hbar} \left(S_{\rm grav}-S_{\rm
grav}^{(flat)}\right)
-{1\over\hbar}S_{\rm em}~,
\eqn\elecN
$$
and
$$
{\rm entropy}=\beta M-\beta F ={1\over \hbar}\left(
S-S^{(flat)}\right)~.
\eqn\elecO
$$
{}From
$$
M={\beta\hbar\over 8\pi G}\left(1-G^2\Phi^4\right)
\eqn\elecP
$$
and
$$
Q={\beta\hbar\over4\pi}\Phi\left(1-G\Phi^2\right)~,
\eqn\elecQ
$$
we then obtain
$$
\beta\hbar=2\pi GM{(1+X)^2\over X}
\eqn\elecP
$$
and
$$
{\rm entropy}=\pi{(GM)^2\over\hbar G}(1+X)^2~,
\eqn\elecR
$$
where we have defined
$$
X\equiv\left(1-{Q^2\over GM^2}\right)^{1/2}
={1-G\Phi^2\over 1+G\Phi^2}~.
\eqn\elecS
$$

\section{The Extreme Solutions}
According to eq.~\elecP, the ``extreme''
Reissner--Nordstr\"om black hole with $Q^2=GM^2$ has
vanishing
temperature.  This result has a simple heuristic
interpretation.  The electrostatic energy $Q^2/2r_+$ stored
in the electric field (outside the event horizon) becomes
comparable to the mass $M$ as the extreme limit is
approached.  Hence there is no mass left over at the center
to support the horizon; the surface gravity approaches zero,
and with it the Hawking temperature.

In fact, as the charge to mass ratio approaches the extreme
value, the inner Cauchy horizon at $r=r_-<r_+$ approaches
the event horizon at $r=r_+$.  The black hole is dangerously
close to becoming a naked singularity.  In this context, the
cosmic censorship hypothesis coincides with the third law of
thermodynamics.  The extreme black hole appears to be a
stable object, unable to shed mass by the Hawking process.

So far, though, we have neglected the coupling of the
electromagnetic field to
electrically charged particles.  (A coupling to particles
with charge
$\hbar e$ was implicit, in the construction of the charge
projection operator.)  The charged particles have no effect
on the leading semiclassical analysis described above, but
have important effects that are higher order in $\hbar$.

The electrostatic potential at the horizon of an extreme
black hole is
$$
\Phi=G^{-1}~,
\eqn\extremeA
$$
(if $Q$ is positive) and so the electrostatic potential
energy of an elementary particle with charge $q$ at the
horizon is $q\Phi=q/G$. In the real world, then, a positron
at the horizon has enormous electrostatic energy compared to
its mass $m$.  Hence, dielectric breakdown of the vacuum
occurs outside the horizon.  It is energetically favorable
to produce an ${\rm e}^+{\rm e}^-$ pair, allowing the
electron to be absorbed by the black hole while the positron
is ejected to spatial infinity with an ultrarelativistic
velocity.  By this process, the black hole neutralizes its
charge and reduces its mass.\Ref\gibcmp{G. W. Gibbons, Comm.
Math. Phys. {\bf 44} (1975) 245.}

As a matter of principle, though, we are free to contemplate
a fictitious world such that the mass $m$ and charge $q$ of
all elementary particles obey
$$
Gm^2>q^2~.
\eqn\extremeB
$$
In this world, the objects with the smallest ratio of mass
to charge are extreme black holes.  Since electric charge is
conserved, the decay of these objects to elementary
particles is kinematically forbidden.

The only decay channel that is potentially available to an
extreme charged black hole is a state that contains black
holes of lower charge and mass.
At the classical level, all extreme black holes have
precisely the same charge--to--mass ratio, so this channel
is just marginally forbidden.  We are obligated, then, to
consider the quantum corrections that might renormalize this
ratio.  There are supersymmetric models in which it is known
that no renormalization occurs; in these models, the extreme
black holes are (just barely)
stable.\Ref\susy{G. W. Gibbons and C. M. Hull, Phys. Lett. {\bf 109B}
(1982) 190.}  Stable extreme
black
holes are fairly generic even in non-supersymmetric models.
The reason is that the charge appearing in the relation
$Q^2=GM^2$ is really the charge of the black hole
renormalized at a length scale comparable to $r_+$, while
the charge that enters the condition for kinematic stability
is the charge renormalized at a large length scale.  If
$r_+$ is small compared to the Compton wavelength $\hbar/m$
(and this is possible for a black hole that is much heavier
than the Planck mass as long as $q^2/\hbar<<1$), then the
charge is more effectively screened by vacuum polarization
for a small black hole than for a large one.  The lighter
extreme black holes, therefore, have more mass per unit
charge than the heavier holes, and the heavier holes
are unable to decay.

These models that contain absolutely stable extreme black
holes are a very intriguing subject for further
investigation.    According to eq.~\elecR, the extreme holes
have a large intrinsic entropy.  Nonvanishing entropy at
zero temperature ordinarily indicates a degenerate ground
state, and so we are challenged to understand the nature of
the degenerate states.  Furthermore, since the extreme black
holes are stable particles, the issue of loss of quantum
coherence comes into sharper focus.  One might hope to
construct an $S$-matrix that describes the scattering of
elementary particles off of an extreme black hole.  But if
such scattering processes inevitably
destroy quantum--mechanical phase information,
then no such $S$-matrix should
exist.

\section{Magnetically Charged Black Hole}
The semiclassical analysis of the thermodynamics of a
magnetically charged black hole is easier than the analysis
of an electrically charged black hole, because the magnetic
charge projection is simpler than the electric charge
projection.

To project out states with magnetic charge $P$, we merely
restrict the path integral to configurations that have
magnetic flux $4\pi P$ on the two-sphere at $r=\infty$.  The
semiclassical evaluation of the partition function in this
sector is dominated by the magnetically charged Euclidean
Reissner--Nordstr\"om solution, which is the analytic
continuation to imaginary time of the Lorentzian
magnetically charged solution.  This Euclidean solution has
the same geometry (and gravitational action) as the
continuation of the electrically charged
Reissner--Nordstr\"om solution,
but it has {\it real} $F_{\theta\phi}$
and {\it positive} electromagnetic action.  In the
semiclassical limit, we find
$$
\beta F(\beta,P)={1\over\hbar}\left(
S_{\rm grav}-S_{\rm grav}^{(flat)}\right)+{1\over
\hbar}S_{\rm em}~.
\eqn\magnetA
$$
Since the analytically continued electrically charged and
magnetically charged solutions have electromagnetic action
of opposite sign, eq.~\magnetA\ is the same function of
$\beta$ and $P$ as the function of $\beta$ and $Q$ in
eq~\elecN.  Thus, the magnetically charged and electrically
charged black holes have identical thermodynamics.  Of
course, this agrees with the standard
analysis;\refmark{\swhextreme} the black
holes with electric charge $Q$ and magnetic charge $P=Q$
have the same surface gravity and surface area.

We find, then, that the Euclidean action of the magnetic
Euclidean Reissner--Nordstr\"om solution is $\beta\hbar~
F(\beta, P)$, where $F$ is the free energy, while the
Euclidean action of the electric Euclidean
Reissner--Nordstr\"om solution is $\beta\hbar
{}~\Omega(\beta,\Phi)$,
where $\Omega$ is the thermodynamic potential.  This agrees
with the analysis of Gibbons and Hawking.

\section{Magnetically Charged Black Holes in Yang--Mills
Theory}
The magnetically charged Reissner--Nordstr\"om black hole is
also a
solution to the field equations of Yang--Mills theory
coupled to
gravity.

Consider, for example, the 't Hooft--Polyakov
model,\refmark{\polyakov} in
which $SO(3)$ is
spontaneously broken to $U(1)$ by the Higgs mechanism.  This
model,
coupled to gravity, actually contains three types of magnetic
monopole solutions,
at least if the symmetry--breaking mass scale is well below
the Planck
mass.

The first (light)
type is essentially the 't Hooft--Polyakov solution,
slightly
perturbed by gravitational effects.  It is non-singular and
has no event
horizon; the $SO(3)$ gauge symmetry is restored in its core.
The second
(heavy)
type is the magnetically charged Reissner--Nordstr\"om solution black
hole.
In this solution,
there is no sign of symmetry restoration near the event
horizon.  The
Higgs field is covariantly constant everywhere outside the
horizon.\Ref\bais{F. A. Bais and R. J. Russell, Phys. Rev. {\bf D11}
(1975) 2692; Y. M. Cho and P. G. O. Freund, Phys. Rev. {\bf D12} (1975)
1588.}

The third type of solution was discovered only very recently.\Ref\Nair{K.
Lee, V. P. Nair, and E. J. Weinberg, ``Black Holes in
Magnetic Monopoles,'' Columbia preprint CU-TP-539 (1991); ``A Classical
Instability of Reissner-Nordstr\"om Solutions and the Fate of
Magnetically Charged Black Holes,'' Columbia preprint CU-TP-540 (1991).}
It may be described as a small black hole embedded inside the core of a
magnetic monopole.  For a sufficiently small black hole, the
Reissner--Nordstr\"om solution is classically unstable.  The solution of
the third type, which has a nontrivial Higgs field core outside the
horizon, is favored instead.  The condition for classical stability of
the Reissner--Nordstr\"om solution is $r_+\ge \sqrt{n}\mu_V^{-1}$, where
$n$ is the magnetic charge in units of $1/e$, and $\mu_V^{-1}$ is the
Compton wavelength of the vector meson that acquires mass via the Higgs
mechanism.
Extreme solutions of the third type do not
exist---black holes inside monopoles always have a finite temperature
and emit Hawking radiation.  The extreme Reissner--Nordstr\"om solution
is classically stable only if $n\ge e^2/(G\mu_V^2)$.

We may also consider a model in which the unbroken gauge
symmetry is
a nonabelian group H.  There will again be both nonsingular
monopoles
and (two types of) magnetically charged black holes.  Suppose, to be
definite, that the
unbroken gauge group is $H=SU(N)/Z_N$; that is, all fields
transform
trivially under the center $Z_N$ of $SU(N)$.
The magnetic field of an $SU(N)$
magnetic monopole can be written in the form
$$
{\bf F}_{\theta\phi}={1\over 2e}{\bf \hat P}\sin\theta~,
\eqn\ymA
$$
where $e$ is the gauge coupling; ${\bf \hat P}$ is a matrix
in the Lie
algebra of $SU(N)$ that, in order to satisfy the Dirac
quantization
condition, must be such that
$$
\exp(2\pi i{\bf\hat P})=e^{2\pi i n/N}{\bf 1}\in Z_N~.
\eqn\ymB
$$
This element of the center is the topological magnetic
charge of the
monopole.\refmark{\erice}

For each nonzero value of the $Z_N$ charge
$n=1,2,3,\dots,N-1$, there is a unique magnetic field
configuration that
is classically stable.\REF\brandt{R. Brandt and F. Neri,
Nucl. Phys. {\bf B161} (1979) 253.}\refmark{\erice,\brandt}
In this configuration (and in a
particular
gauge), the matrix ${\bf \hat P}$ is the diagonal matrix
$$
{\bf \hat P}={\bf \hat P_n}\equiv
{\rm diag}\left(\underbrace{{n\over N},\dots,{n\over N}}_
{N-n~~{\rm times}}~,~
\underbrace{{n-N\over N},\dots,{n-N\over N}}_
{n~~ {\rm times}}\right)~.
\eqn\ymC
$$

Of course, if $SU(N)$ is unbroken, then we expect this
theory to be
confining, and the classical magnetic field will be screened
at long range.
We will further consider the consequences of this screening
in
Section~5.  For now, we merely note that, since the $Z_N$
magnetic
charge has a
topological meaning, it is easy to construct a projection
onto a
sector with a definite value of the $Z_N$ charge, by
restricting the
path integral to field configurations with a specified $Z_N$
flux on the
two-sphere at $r=\infty$.  The black hole
partition function in each nontrivial
sector, in the semiclassical limit, is dominated by the stable classical
black hole solution---either the
Euclidean Reissner--Nordstr\"om solution, with the magnetic
field
given by eq.~\ymA\ and eq.~\ymC, or, if the
Reissner--Nordstr\"om  solution is unstable, a solution analogous to
that described in Ref.~\Nair.  (But effects that are {\it
nonperturbative} in $\hbar$ can drastically change the
story, as we will
see in Section~5.)

To complete this classical discussion, we will find the mass
of the
extreme Reissner--Nordstr\"om black hole with $Z_N$ charge $n$.
The easiest way
to do this is
to apply the result that the extreme black holes in the
abelian theory
have $GM^2=P^2$; acting as a source in the Einstein
equations, the field
of an $SU(N)$ monopole with charge $n$ is equivalent to that
of a $U(1)$
monopole with a charge $P_n$ that we can
compute.\Ref\gupta{A. K.
Gupta, J. Hughes, J. Preskill, and M. B. Wise, Nucl. Phys.
{\bf B333} (1990) 195.}  The
conventional
normalization of the $SU(N)$ generators (and of the gauge
coupling $e$)\foot{But see the remark about the
normalization of $e$ in Section~4.2 below.}
is such that ${\rm tr}~{\bf F}_{\theta\phi}^2$
in the nonabelian theory is equivalent
to ${1\over 2}F_{\theta\phi}^2$ in the abelian theory.
Hence, the mass
$M_n$ of the extreme black hole with $Z_N$ charge $n$ is
given by
$$
GM_n^2=P_n^2\equiv {1\over 2e^2}{\rm tr}~{\bf \hat P}^2
=\left({1\over 2e^2}\right){n(N-n)\over
N}~,~~n=1,2,\dots,N-1~.
\eqn\ymD
$$
If the dimensionless coupling constant $\hbar e^2$
(renormalized at a
distance scale comparable to the size of the black hole) is
small, then
$M_n$ is much larger than the Planck mass, so that
semiclassical methods
are reliable.

If there are light nonsingular monopoles that carry the
$Z_N$ charge,
then the extreme Reissner--Nordstr\"om solution will be classically
unstable.\refmark{\Nair}
But if the symmetry breaking mass scale is large
enough,
then the ($n=\pm 1$) magnetically charged
Reissner--Nordstr\"om black hole is the
lightest
magnetically charged object in the theory,
and must be absolutely stable.  In fact, in
that event, the magnetically charged black hole is stable
for each value
of the charge $n$,
for  according to eq.~\ymD, there is no lighter object with
the same
total magnetic charge.

\chapter{Screened Electric Charge}
\section{$Z_N$ Electric Charge Projection}
We will now consider an abelian Higgs model in which $U(1)$
is
spontaneously broken to $Z_N$, as described in Section~2.
$Z_N$ electric
charge is screened, and so, according to the no--hair
theorem, the $Z_N$
electric charge on a stationary black hole has no effect on
the geometry of
the hole, in the classical approximation.  But the charge
does have quantum
effects, effects that, as we will see, are nonperturbative
in $\hbar$.  We
wish to calculate how the $Z_N$ electric charge modifies the
thermodynamic
behavior of the black hole.\refmark{\cpwA,\cpwB}

In this model, the charge superselection sectors are labeled
by the $Z_N$
charge, and our first task is to find the Euclidean path
integral
prescription for computing the partition function restricted
to a
particular charge sector.  For this purpose, we note that
the periodic
configurations of finite action must be such that $$
e\int_0^{\beta\hbar}d\tau~A_\tau(\tau,\vec
x)\big|_{r=\infty}= {2\pi\over
N} k~,
\eqn\NprojA
$$ where $k$ is an integer.  (Otherwise, the contribution to
the action
arising from the covariant derivative of the Higgs field is
divergent.)
Thus, the integral over $\omega$ in eq.~\projM\ collapses to
a sum over
$k$.  We therefore find that $$ Z(\beta,Q)={1\over
N}\sum_{k=-\infty}^{\infty} e^{-2\pi i k Q/N\hbar
e}Z(\beta,k)~,
\eqn\NprojB
$$ where $$ Z(\beta,k)=\int_{\beta\hbar,k}e^{-
S_E/\hbar}\eqn\NprojC $$ is
the Euclidean path integral over configurations that are
periodic in Euclidean time with period
$\beta\hbar$ and satisfy the constraint
eq.~\NprojA.  Evidently,
$Z(\beta,Q)$ depends only on $Q$ modulo $N\hbar e$, the
charge that is not
screened by the condensate.

An alternative way to derive eq.~\NprojB\ is to repeat the
derivation in
Section~3.3, but inserting the $Z_N$ charge--projection
operator $$
P_Q={1\over N}\sum_{k=0}^{N-1}e^{-2\pi i k Q/N\hbar e}U(k)
\eqn\NprojD
$$ into the path integral, where $U(k)$, the operator that
represents a
local $Z_N$ transformation, acts on a charge-$Q$ state
according to $$
U(k)\ket{Q}=e^{2\pi i k Q/N\hbar e}\ket{Q}~.
\eqn\NprojE
$$

If we are interested in the case of a black hole with
specified $Z_N$
electric charge, then we integrate over the geometry as
well, restricted to
the sector with topology $R^2\times S^2$.  In this context,
eq.~\NprojA\
has a rather remarkable interpretation.  If we regard
$F_{r\tau}$ as a {\it
magnetic} field, then
$\int_0^{\beta\hbar}d\tau~A_\tau=\int dr~d\tau~F_{r\tau}$
is
the magnetic
flux in the $r$--$\tau$ plane, and $k$ is the {\it
vorticity}, the value of
the flux in units of the flux quantum $2\pi/Ne$.  The
electric charge $Q$
determines the phases that weight the different vorticity
sectors.  In
fact, as we will explain in more detail below, the integer
$k$ can be
interpreted as the net number of virtual cosmic string world
sheets that
wrap around the black hole.  The phase that accompanies the
contribution
from the sector with vorticity $k$ in eq.~\NprojB, then, is
precisely the
Aharonov--Bohm phase acquired by the virtual string, if the
black hole
carries charge $Q$.

We will briefly explain how the results of this section can
be generalized to the case in which the unbroken local
symmetry group is a {\it nonabelian} finite group $H$.  In
this case, an operator can be constructed that projects out
states that transform as a specified irreducible
representation $(\mu)$ of $H$; this projection operator is
$$
P_{(\mu)}={n_\mu\over n_H}\sum_{h\in H}
\chi^{(\mu)}(h)^*~U(h)~.
\eqn\NprojF
$$
Here $U(h)$ is the operator that represents $h\in H$ acting
on Hilbert space, $n_H$ is the order of the group,
$\chi^{(\mu)}$ is the character of the irreducible
representation $(\mu)$, and $n_\mu$ is the dimension of
$(\mu)$.  By inserting this projection operator into the
path integral, we obtain an expression for the partition
function restricted to the charge sector containing states
that transform according to $(\mu)$.  The result is
$$
Z(\beta,(\mu))={n_\mu^2\over n_H}\sum_{h\in H}
{1\over n_\mu}\chi^{(\mu)}(h)^*~Z(\beta,h)~,
\eqn\NprojG
$$
where $Z(\beta,h)$ is the Euclidean path integral over
configurations that are periodic in Euclidean time with
period $\beta\hbar$ and satisfy the constraint
$$
P\exp\left(ig\int_0^{\beta\hbar}d\tau
A_\tau(\tau,\vec x)\big|_{r=\infty}\right)=h~.
\eqn\NprojH
$$

For the case of a black hole with $H$-charge, the constraint
requires a vortex with ``flux'' $h\in H$ to occupy the
$r$-$\tau$ plane.  Once again, the interpretation is clear.
The
factor $(1/n_\mu)\chi^{(\mu)}(h)^*$ that weights the sector
with vortex flux $h$ is precisely the Aharonov--Bohm factor
acquired by a virtual cosmic string with flux $h$ that winds
around the black hole.  The reason that this ``phase'' has
modulus less than one is that a string that is initially
uncharged may exchange charge with the black hole during the
winding, and the virtual string can re-annihilate only if it
remains uncharged.  The probability that the string remains
uncharged after winding around an object that carries charge
$(\mu)$ is just $|(1/n_\mu)\chi^{(\mu)}(h)|^2$, the absolute
value squared of the weight factor appearing in
eq.~\NprojG.\Ref\bucher{M. Bucher, K.-M. Lee, and J.
Preskill,
``On Detecting Discrete Cheshire Charge,'' Caltech preprint
CALT-68-1753 (1991).}

\section{Semiclassical Thermodynamics}
As in Section~3, we find the semiclassical thermodynamic
behavior of a
black hole with $Z_N$ electric charge by evaluating
eq.~\NprojB\ in the
limit $\hbar\to 0$, with $\beta\hbar$ held fixed.  But now,
in taking the
limit, we will treat the charge $Q$ differently than before.

In the standard semiclassical limit (invoked in Section~3),
the classical
charge $Q$ is held fixed as $\hbar$ gets small.  In this
limit, the charge
quantum $\hbar e$ becomes arbitrarily small compared to $Q$.
And the
Aharonov--Bohm phase $\exp(-2\pi i Q/N\hbar e)$, acquired
when charge $Q$
circumnavigates flux $2\pi/Ne$, oscillates wildly.  The
oscillations wipe
out the Aharonov--Bohm interference; the Aharonov--Bohm
effect (which is
quantum mechanical) does not survive in the standard
classical limit.

We are interested in a different case, in which the $Z_N$
charge is held
fixed as $\hbar$ gets small.  Thus, we fix $Q/N\hbar e$, the
classical
charge in units of the charge of the condensate, rather than
$Q$.  Then the
Aharonov--Bohm phases appearing in eq.~\NprojB\ are
well--behaved, and
effects that depend on the $Z_N$ charge can be investigated.

The model that we are studying has the action
$$ S=S_{\rm grav}+S_{\rm
em}+S_{\rm Higgs}~,
\eqn\semithermA
$$
with $S_{\rm grav}$ as in eq.~\schwarzG, $S_{\rm em}$ as in
eq.~\elecB,
and \foot{In the subsequent discussion, we set $\xi=0$.  The
results are qualitatively similar for $\xi\ne 0$.}
$$ \eqalign{
S_{\rm Higgs}={1\over 4\pi}\int d^4x\sqrt{g} &
\biggl[ g^{\mu\nu} \left(\partial_\mu-ieNA_\mu\right)\phi^*
\left(\partial_\nu+ieNA_\nu\right)\phi \cr
&+{\lambda\over 2}\left(|\phi|^2-{v^2\over 2}\right)^2
+\xi|\phi|^2 R
\biggr]~.\cr}
\eqn\semithermB
$$
(We may ignore the fields that carry nontrivial $Z_N$
electric charge,
for they have no effect on the leading semiclassical
result.)
Note that we have adopted (with some reluctance)
unrationalized units
such that the gauge field kinetic term in the Euclidean
Lagrange density is
$(1/8\pi)(E^2+B^2)$.  We do so to be consistent with the
notation of
Section~3, and with most of the literature on charged black
holes.  (In the
rationalized units usually favored by particle physicists,
the gauge field
kinetic term is $(1/2)(E^2+B^2)$, and the condition
satisfied by an extreme
Reissner--Nordstr\"om black hole is $Q^2=4\pi GM^2$.)  Thus
we are
normalizing the gauge coupling $e$ in a manner than departs
from the common
convention; the (more common) rationalized coupling $e_{\rm
rat}$ is
related to our $e$ by $$ e^2={e_{\rm rat}^2\over 4\pi}
\eqn\semithermC
$$ We wish to caution the reader that, in our units, the
``fine structure
constant'' is $\alpha=\hbar e^2=\hbar e_{\rm rat}^2/4\pi$.

To evaluate $Z(\beta,k)$ in the semiclassical limit, we seek
the solution
to the Euclidean field equations, of lowest action, that has
topology
$R^2\times S^2$, is asymptotically flat, is periodic in
$\tau$ with period
$\beta\hbar$, and has vorticity $k$.  Since the magnetic
flux quantum is
independent of $\hbar$, increasing the vorticity increases
the action of
this solution by an amount that is independent of $\hbar$.
Therefore, to
compute the leading charge--dependent corrections to the
partition function
in the semiclassical limit, we need only retain the sectors
with vorticity
$k=0$ and $k=\pm 1$.

Hence we find $$ Z(\beta,Q)\simeq {1\over
N}\left[Z(\beta,k=0)+2\cos
\left({2\pi Q\over N\hbar e}\right)Z(\beta,k=1)\right]~,
\eqn\semithermDD
$$ or $$ {Z(\beta,Q)\over Z(\beta,Q=0)}\simeq 1-2\left[1-
\cos\left({2\pi
Q\over N\hbar e}\right)\right] {Z(\beta,k=1)\over
Z(\beta,k=0)}~.
\eqn\semithermD
$$ The $k=0$ solution is just the Euclidean Schwarzschild
solution, so we
have $$ {Z(\beta,k=1)\over Z(\beta,k=0)}\simeq{1\over
2}C(\beta\hbar)
e^{-\Delta S_{\rm vortex}/\hbar}~.
\eqn\semithermE
$$ Here, $$
\Delta S_{\rm vortex}= S_{\rm vortex}-S_{\rm
Schwarzschild}~,
\eqn\semithermF
$$ where $S_{\rm vortex}$ is the action of the $k=1$
solution, and
$(1/2)C(\beta\hbar)>0$ is a ratio of functional
determinants.  (This ratio is positive because the
phases that arise from the negative modes in the integration over
geometries
are divided out.)
Taking the
logarithm of eq.~\semithermD, we find the
leading charge--dependent
contribution to the black hole free energy $$
\beta F(\beta,Q)-\beta F(\beta,Q=0)\simeq
\left[1-\cos\left({2\pi Q\over N\hbar e}\right)\right]
C(\beta\hbar)e^{-\Delta S_{\rm vortex}/\hbar}~.
\eqn\semithermG
$$

Now we may invoke the thermodynamic identity $$
M(\beta,Q)=\partder{(\beta
F)}{\beta}
\eqn\semithermH
$$ to find $$ M(\beta,Q)-M(\beta,Q=0)\simeq
-\left[1-\cos\left({2\pi Q\over N\hbar e}\right)\right]~
\partder{\Delta S_{\rm vortex}}{(\beta\hbar)}~
C(\beta\hbar)e^{-\Delta S_{\rm vortex}/\hbar}~,
\eqn\semithermI
$$ the expression relating the mass, charge, and
temperature.  (We have
neglected a term that is suppressed by an additional power
of $\hbar$.)

Note the significance of the sign of $\partial\Delta S_{\rm
vortex}/\partial(\beta\hbar)$.  If it is positive, then
adding charge reduces the mass of a black hole of given
temperature, and correspondingly reduces the temperature of
a black hole of given mass.  Conversely, if
$\partial\Delta S_{\rm vortex}/\partial(\beta\hbar)$ is
negative, then adding charge heats up a black hole of given
mass.

To go further, we must calculate $\Delta S_{\rm vortex}$.
Let us assume that the vortex is rotationally invariant,
satisfying the {\it Ansatz}
$$
\phi={\rho(r)\over \sqrt{2}}\exp(-2\pi i \tau/\beta\hbar)~,
\eqn\semithermJ
$$
$$
eA_\tau={2\pi\over N\beta\hbar}[1-a(r)]~,
\eqn\semithermK
$$
and with the geometry of the form eq.~\schwarzB.  Then the
action can be expressed as
$$
\eqalign{
S_{\rm grav}-S_{\rm grav}^{(flat)}=&
\beta\hbar\int_{r_+}^{\infty}dr\left(-{1\over 2G}\right)
\left[e^{(\Phi+\Lambda)}+e^{(\Phi-\Lambda)}(2r\Lambda'-
1)\right]\cr
&-{1\over G}\pi r_+^2+{\beta\hbar\over G}~
\lim_{r\to \infty}\left[re^\Phi (1-e^{-\Lambda})\right]}
\eqn\semithermL
$$
(where we have done an integration by parts, and used
eq.~\schwarzD\ and eq.~\schwarzJ - \schwarzK), and
$$
\eqalign{
S_{\rm matter}\equiv S_{\rm em}+S_{\rm Higgs}&\cr
=\beta\hbar\int_{r_+}^{\infty}dr~ r^2
\biggl\lbrace&{1\over 2}\left({2\pi\over
Ne\beta\hbar}\right)^2
\left[e^{-(\Phi+\Lambda)}(a')^2+e^{(\Lambda-
\Phi)}(Ne)^2\rho^2a^2\right]
\cr
&+e^{(\Phi-\Lambda)}{1\over 2}(\rho')^2+
e^{(\Phi+\Lambda)}{\lambda\over 8}
\left(\rho^2-v^2\right)^2\biggr\rbrace~.}
\eqn\semithermM
$$
(The primes denote derivatives with respect to $r$.)  By
demanding that the action is stationary, we obtain coupled
differential equations satisfied by $\Phi(r)$, $\Lambda(r)$,
$\rho(r)$, and $a(r)$.

In the next two sections, we will describe how these
equations can be solved
analytically, and the action of the solution explicitly
computed, in two different
limiting cases.  First, though, we remark on how
eq.~\semithermD\ generalizes to the case in which the
unbroken local symmetry group is a (not necessarily abelian)
discrete group $H$.  In eq.~\NprojG, we expressed the
partition function $Z(\beta,(\mu))$ of a black hole with
$H$-charge $(\mu)$ as a weighted sum of contributions from
sectors of specified vorticity.  The path integral
$Z(\beta,h)$ restricted to the sector with vorticity $h$
actually depends only on the conjugacy class to which $h$
belongs, for $h\to h'hh'^{-1}$ under a global $H$
transformation.  Furthermore, $CPT$-invariance implies
$Z(\beta,h)=Z(\beta,h^{-1})$; an $h$ string that wraps in
one sense around the black hole is equivalent to an $h^{-1}$
string that wraps in the opposite sense.  Barring further
symmetries or accidental degeneracies, the charge dependence
of $Z(\beta,(\mu))$ will be dominated in the semiclassical
limit by vortices belonging to a single class (and the
inverse class), the vortices of lowest Euclidean action.  If
we denote the class that dominates by $\alpha$, then the
generalization of eq.~\semithermD\ becomes
$$
{Z(\beta,(\mu))\over Z(\beta,(\mu)=(0))}\simeq
n_\mu^2\left(1-2n_\alpha\left[1-{1\over n_\mu}
{\rm Re}~ \chi^{(\mu)}_\alpha\right]{Z(\beta,\alpha)\over
Z(\beta,\{e\})}\right)~,
\eqn\semithermN
$$
where $n_\alpha$ is the order of the class.

\section{The Thin String Limit}
We first consider the limit in which the natural thickness
$\mu^{-1}$ of
the string is much less than the radius $r_+\simeq 2GM$ of
the event
horizon of the black hole. (Here $\mu^{-1}=(Nev)^{-1}$ is
the Compton
wavelength of the massive vector meson.)  In this limit, the
vortex shrinks
to a point at the origin of the $r$--$\tau$ plane.  The
matter field
configuration near the origin looks just like the cross
section of a
straight cosmic string in flat space.  We may interpret this
configuration
as the world sheet of a virtual string wrapped tightly
around the minimal
two--sphere (the ``horizon'') of the Euclidean Schwarzschild
geometry.

To start with, we will make an additional assumption, that
the string
tension is small in Planck units, or
$$
G{\tilde T}_{\rm string}<<1~.
\eqn\thina
$$
Here ${\tilde T}_{\rm string}$ is the tension of a straight
cosmic
string in
flat space; it has the form
$$
{\tilde T}_{\rm string}= {1\over 4}v^2~f(\lambda/e^2)~,
\eqn\thinAA
$$
where $f$ is a slowly varying function such that
$f(1)=1$.\Ref\Bogo{E.
B. Bogomol'nyi, Sov. J. Nucl. Phys. {\bf 24} (1976) 449.}
(There is a gravitational correction to eq.~\thinAA, as the
string has a
$G$-dependent gravitational self--energy; this correction
may be ignored
under the assumption eq.~\thina.)

Now there are two contributions to the action that we must
consider---the
contribution $S_{\rm matter}=S_{\rm em}+S_{\rm Higgs}$ due
to the matter fields, and the
gravitational contribution $S_{\rm grav}$.  Since the matter
field
configuration is a vortex
localized on the minimal two--sphere, we have
$$
S_{\rm matter}
=4\pi r_+^2{\tilde T}_{\rm string} ~.
\eqn\thinA
$$
To estimate $S_{\rm grav}$, we note that, for $G{\tilde
T}_{\rm string}<<1$,
the geometry of the vortex
solution is close to the Euclidean Schwarzschild geometry;
the perturbation of the geometry is of linear order in
$G{\tilde T}_{\rm
string}$.  And since the Euclidean Schwarzschild geometry is
a stationary
point of the Einstein--Hilbert action, the change in  the
gravitational action is of second order.  Thus, we have
$$
\Delta S_{\rm grav} = {1\over 16 \pi G}~
o\left[(G{\tilde T}_{\rm string})^2\right]~,
\eqn\thinaa
$$
which can be neglected compared to $S_{\rm matter}$.  We
therefore find
$$
\Delta S_{\rm vortex}\simeq {(\beta\hbar)^2\over 4\pi}
{\tilde T}_{\rm string}~,~~~G{\tilde T}_{\rm string}<<1~.
\eqn\thinaaa
$$

In fact, we can go further, and find an expression for
$\Delta S_{\rm
vortex}$ in the thin string limit without making the
assumption
eq.~\thina.  We may replace
$S_{\rm matter}$  by an effective action---the Nambu-Goto
action for a
relativistic string with tension $T_{\rm string}$; thus the
``matter''
action becomes
$$
S_{\rm NG}= ({\rm World~Sheet~Area})~T_{\rm string}~.
\eqn\thinb
$$
Here $T_{\rm
string}$ is not, in general, the same as ${\tilde T}_{\rm
string}$
given by eq.~\thinAA, because $T_{\rm string}$ may include
$G$-dependent
gravitational corrections.
In the vortex solution, eq.~\thinb\ coincides with
eq.~\thinA, except that
$T_{\rm string}$ has replaced ${\tilde T}_{\rm string}$.

Now consider $S_{\rm grav}$.
The
stress tensor of a static straight string along the $z$
axis,
integrated over a planar slice perpendicular to the string,
has the form
$$
\int d^2 x\sqrt{h}~\theta^\mu_{~\nu}=
T_{\rm string}~{\rm diag}(1,0,0,1)~
\eqn\thinaaaa
$$
(where $h$ is the induced metric on the slice).
{}From the Einstein equation, we have
$$
R=8\pi G~\theta^\mu_{~\mu}~,
\eqn\thinbb
$$
and therefore, using eq.~\thinaaaa,
$$
-~{1\over 16\pi G}\int d^4x\sqrt{g}R=-~({\rm Area})~T_{\rm
string}~,
\eqn\thinbbb
$$
which exactly cancels eq.~\thinb.

It remains to compute $S_{\rm boundary}$.  Outside of the
pointlike
vortex, the
geometry is
(locally) a Euclidean Schwarzschild
solution.  But the
point vortex can modify the relation between $\beta\hbar$
and $S_{\rm
boundary}$.  To find the correct relation, it is helpful to
notice that the
problem of finding the geometry outside the vortex, but at a
distance from
the vortex that is much less than $r_+$, is equivalent to
another problem
with a well--known solution---the problem of finding the
geometry outside a
static straight string.\Ref\vilenkin{A. Vilenkin, Phys. Rev. {\bf D23}
(1981) 852.}  The geometry outside the static
straight string is
a conical space with a deficit angle $$
\delta=8\pi G T_{\rm string} ~.
\eqn\thinB
$$
It follows that the effect of the point vortex at the
origin is to
modify the condition eq.~\schwarzD; it is replaced by $$
2\pi -
\delta={\beta\hbar\over 4GM}~,
\eqn\thinC
$$ where $M$ is the mass appearing in the Euclidean
Schwarzschild solution
that describes the geometry outside the vortex.  Now, the
boundary
contribution to the action of the vortex solution is related
to $M$ as in
eq.~\schwarzN, so we find $$ S_{\rm vortex}-S_{\rm
grav}^{(flat)}={(\beta\hbar)^2\over 16\pi G}
\left(1-4GT_{\rm string}\right)^{-1}~,
\eqn\thinD
$$
or
$$
\Delta S_{\rm vortex}={1\over 4\pi}(\beta\hbar)^2T_{\rm
string}~
 (1-4GT_{\rm string})^{-1}~.
\eqn\thinE
$$

By combining eq.~\thinE\ and eq.~\semithermI, we find the
leading
charge--dependent contribution to the mass of a black hole
of given
temperature.  Since we also know that the leading
charge--independent
contribution to the mass is $M\simeq  \beta\hbar/8\pi G$, we
can invert
eq.~\semithermI, and find the charge--dependent correction
to the
temperature of a black hole of given mass.  The result is $$
\eqalign{
\beta^{-1}(M,Q)-\beta^{-1}(M,Q=0)\simeq &\cr
-~{\hbar \eta T_{\rm string}\over 2\pi M}~C(8\pi GM)~
\left[1-\cos\left({2\pi Q\over N\hbar e}\right)\right]~&
\exp\left(-16\eta\pi (GM)^2T_{\rm string}/\hbar\right)~,\cr}
\eqn\thinF
$$ where we have defined $$ \eta= (1-
4GT_{\rm
string})^{-1}~.
\eqn\thinG
$$ Thus, the $Z_N$ electric charge on a black hole lowers
its temperature
compared to the temperature of an uncharged hole with the
same mass.  This
effect has the same sign as the effect of unscreened
electric charge, in
the case of the Reissner--Nordstr\"om black hole.

\section{The Thick String Limit}
We now consider the opposite limit, in which the natural
string thickness is large compared to the size of the black
hole.  In this limit, the action of the vortex can be
expanded in powers of $(\mu r_+)^2$.  The leading term in
this expansion can be obtained by setting $v^2=0$ in
eq.~\semithermM.  Then the action is minimized by setting
$\rho=0$ everywhere, and the equations to be solved become
identical to those for the purely electromagnetic case,
which we have already analyzed in Section~3.5.  Thus, the
vortex action is given by eq.~\elecH, with $\omega=2\pi/N$.
We therefore have
$$
\Delta S_{\rm vortex}={\pi\over 2(Ne)^2}
\left[1+{1\over 2}{G\over(\beta\hbar)^2}
\left({2\pi\over Ne}\right)^2\right]~.
\eqn\thickA
$$

In the limit $\hbar\to 0$ and $v^2\to 0$, the leading
charge--dependent contribution to the mass of a black hole
of given temperature is found by combining eq.~\thickA\ with
eq.~\semithermI.  We may invert this relation, as before, to
find the charge--dependent correction to the temperature of
a black hole with given mass.  The result is
$$
\eqalign{
\beta^{-1}(M,Q)&-\beta^{-1}(M,Q=0)\simeq\cr
{\hbar\over 2048\pi G^3M^5(Ne)^4}~& C(8\pi GM)
\left[1-\cos\left({2\pi Q\over N\hbar e}\right)\right]\cr
&\times\exp\biggl\lbrace -{\pi\over 2(Ne)^2}\left[1+{1\over
32GM^2(Ne)^2}\right]\biggr\rbrace~.}
\eqn\thickB
$$
We see that, in the thick string limit, adding charge to a
black hole of fixed mass actually causes the black hole to
heat up.  This behavior is the opposite of that found in the
thin string limit or in the Reissner--Nordstr\"om case.

The leading $v^2$--dependent correction to $\Delta S_{\rm
vortex}$ is of order $v^2r_+^2=(1/(Ne)^2)(\mu r_+)^2$, and
so is small compared to the expression in eq.~\thickA\
provided $\mu r_+<<1$, which is just the condition for the
natural string thickness to be much larger than the size of
the black hole.  It is instructive to compare eq.~\thickA\
with eq.~\thinE, which was derived under the assumption $\mu
r_+>>1$.  Suppose that the back reaction of the matter
fields on the geometry is a small effect, so that the
background geometry of the vortex is very close to the
Euclidean Schwarzschild geometry; this is true if
$$
GT_{\rm string}<<1
\eqn\thickC
$$
in the thin string limit, or if
$$
{G\over (\beta\hbar)^2}\left({2\pi \over Ne}\right)^2<<1
\eqn\thickD
$$
in the thick string limit.  We then have
$$
\Delta S_{\rm vortex}^{(thin)}\sim 2\pi v^2 r_+^2
\eqn\thickE
$$
(taking $T_{\rm string}\sim (1/4)v^2$),
and
$$
\Delta S_{\rm vortex}^{(thick)}\simeq {\pi\over 2(Ne)^2}~.
\eqn\thickF
$$
Thus, the thin--string and thick--string expressions for the
vortex action
cross when $r_+\sim\mu^{-1}=(Nev)^{-1}$, as one would
expect.

We note that eq.~\thickF\ has a simple interpretation.
Recall that the vortex may be interpreted as a virtual
string world sheet that envelops the black hole.  Because
the string has a finite thickness of order $\mu^{-1}$, the
action of a string world sheet has a nonzero minimum value.
In flat space, this minimum is of order
$$
S_{\rm world~sheet}\sim 4\pi \mu^{-2}T_{\rm string}\sim
{\pi\over (Ne)^2}~.
\eqn\thickG
$$
This agrees reasonably well with eq.~\thickF.

\section{The Electric Field}
We emphasized in Section~2 that the $Z_N$ electric charge on
a black hole can be measured by means of the Aharonov--Bohm
interaction of the hole with a cosmic string.  Now we have
seen that an observer who is not equipped with a cosmic
string can also detect the charge, by measuring both the
mass and the temperature of the black hole radiation.

Even this is not the whole story.  Although the black hole
has no {\it classical} hair, various local observables
acquire charge--dependent expectation values due to quantum
effects.  In particular, there is an electric field outside
the event horizon; it could be detected by an experimenter
armed with electroscopes and pith balls.

To understand the origin of this electric field, consider
the virtual process in which a loop of cosmic string
nucleates at a point on the event horizon of a black hole,
sweeps around the horizon two--sphere, then shrinks and
annihilates at the antipodal point.  The virtual string has
magnetic flux in its core; hence its motion creates an
electric field orthogonal to the magnetic field and the
direction of motion, an electric field in the radial
direction.  The time--integrated value of this radial field
is purely geometrical---although the electric field is
proportional to velocity, the time that the string spends at
any point on the sphere is inversely proportional to
velocity.

We must also average over all possible points of nucleation.
This averaging cancels the magnetic field of the string, but
the radial electric field survives.
We now recognize that our vortex solution, which is
rotationally invariant and has nonzero $F_{r\tau}$,
represents this averaged string world sheet.

Now we add together the contributions from the sectors with
vorticity $k=\pm 1$, which correspond to the two possible
orientations of the virtual string.  String world sheets of
opposite orientation generate electric fields of opposite
sign; their contributions would cancel if the two sectors
were weighted equally.  But if the $Z_N$ charge on the black
hole is nonzero, then the $k=\pm 1$ sectors are weighted by
unequal Aharonov--Bohm phases, and the electric field
acquires a non-trivial expectation value.

The calculation of the electric field is greatly simplified
if we assume that eq.~\thickC\ and eq.~\thickD\ are
satisfied; then it is a good approximation to neglect the
back reaction of the vortex on the geometry.  Because the
solutions with $k=0$ and $k=\pm 1$ have the same geometry,
we can easily compute the expectation value of a local
observable at a fixed spacetime point in the vicinity of the
black hole.  If we do not make this approximation, then the
computation of expectation values of observables is a much
more complicated and delicate task.

In each charge sector, we compute the (Euclidean)
expectation value of an observable ${\cal O}$ as
$$
\VEV{\cal O}_{\beta,Q}^{({\rm Euc})}={1\over Z(\beta,Q)}
{}~{1\over N}\sum_{k=-\infty}^{\infty}e^{-2\pi i k Q/N\hbar
e}
\int_{\beta\hbar,k}{\cal O}~e^{-S_E/\hbar}~,
\eqn\fieldA
$$
where the path integral is restricted to configurations with
vorticity $k$, and $Z(\beta,Q)$ denotes the partition
function in the sector with charge $Q$.
Lorentzian expectation values may
then be obtained by continuing to real time.

Applying this formula to the electric field in the black
hole sector, we find, in the semiclassical limit,
$$
\VEV{F_{r\tau}}_{\beta,Q}^{\rm (Euc)}=-i
\sin\left({2\pi Q\over N\hbar e}\right)
C(\beta\hbar)e^{-\Delta S_{\rm vortex}/\hbar}
\left(F_{r\tau}\right)_{\rm vortex}~,
\eqn\fieldB
$$
where $(F)_{\rm vortex}$ denotes $F$ in the $k=1$ vortex
solution.  By continuing to real time, we find the
expectation value of the radial electric field,
$$
\VEV{E(r)}_{\beta,Q}=\sin\left({2\pi Q\over N\hbar e}\right)
C(\beta\hbar)e^{-\Delta S_{\rm vortex}/\hbar}
\left(F_{r\tau}(r)\right)_{\rm vortex}~.
\eqn\fieldC
$$

If we neglect back reaction, then we may find $(F)_{\rm
vortex}$ by solving the matter field equations on the
Euclidean Schwarzschild background geometry.
In the thick string limit, the electric field of the vortex
is well--approximated, for $r<<\mu^{-1}$, by the field of
the Euclidean Reissner--Nordstr\"om solution.  We therefore
have
$$
\left(F_{r\tau}(r)\right)_{\rm vortex}\simeq
\left({1\over 2Ne}\right){1\over r^2}~,~~~r<<\mu^{-1}~.
\eqn\fieldD
$$
We can also find the asymptotic large-$r$ behavior of the
electric field by solving the field equations perturbatively
in $1/r$, without making any assumption about whether $\mu
(2GM)$ is large or small.  The result is
$$
\left(F_{r\tau}(r)\right)_{\rm vortex}\simeq
C~\left({1\over 2Ne}\right)
\left({\mu\over r}+{1\over r^2}\right)
\left({2GM\over r}\right)^{\mu GM}~e^{-\mu r}~,~~~r>>\mu^{-
1},GM~,
\eqn\fieldE
$$
where $C$ is a numerical constant of order one, and
corrections down by further powers of $GM/r$ have been
neglected. Thus we have found that the expectation value of
the electric field strength outside the horizon is
nonvanishing for $Q\ne 0$, and that the field decays
exponentially far away from the black hole.

Similarly, we can compute, in the leading semiclassical
approximation, the expectation values of all powers of the
electric field, obtaining
$$
\eqalign{
&\VEV{E(r)^n}_{\beta,Q}=~~{\rm charge-independent}\cr
&~+~(-1)^{n/2}\cos\left({2\pi Q\over N\hbar e}\right)
C(\beta\hbar)e^{-\Delta S_{\rm vortex}/\hbar}
\left(F_{r\tau}(r)^n\right)_{\rm vortex}~~~(n~{\rm
even})~,\cr}
\eqn\fieldC
$$
$$
\eqalign{
&\VEV{E(r)^n}_{\beta,Q}=\cr
&~(-1)^{(n-1)/2}
\sin\left({2\pi Q\over N\hbar e}\right)
C(\beta\hbar)e^{-\Delta S_{\rm vortex}/\hbar}
\left(F_{r\tau}(r)^n\right)_{\rm vortex}~~~(n~{\rm
odd})~.\cr}
\eqn\fieldD
$$
We emphasize that the leading contributions to the expectation values
of even powers of $E$ are charge-independent. They are just the effects of the
ordinary vacuum fluctuations of the free electromagnetic field (in a
gravitational background), or, expressed in path-integral language, the
effects of one-loop corrections in the $k=0$ sector.  The charge-dependent
corrections to these are exponentially small additions; thus there is no
need for them to be nonnegative, and they are not.

Eqs. \fieldC\ and \fieldD\ do not describe the moments of a probability
distribution in which an exponentially small background field is added to the
usual distribution of vacuum fluctuations. Rather, exponentially rare events
(in no one of which the field is extraordinarily small) are added to the usual
distribution. This is typical of tunneling processes.  Consider the
measurement of the total energy in some region near an alpha-unstable nucleus.
Most of the time, all one detects are vacuum fluctuations, but every once in
an exponentially rare while an alpha particle comes by.

If we attempt to construct $\rho(E)$, the probability distribution whose
moments match the expectation values \fieldC\ and \fieldD, we encounter an
apparent paradox.  The obvious answer is
$$\rho(E) = {\rm charge-independent} + a\delta(E-E_0) + a^*\delta(E-E_0^*)~,
\eqn\fieldE
$$
where
$$ a={\textstyle {1 \over 2}}C(\beta\hbar)e^{-\Delta S_{\rm vortex}/\hbar}
e^{-2\pi iQ/N\hbar e},\qquad {\rm and}\qquad
E_0=i(F_{r\tau}(r))_{\rm vortex}~.
\eqn\fieldF
$$
But this is preposterous; a probability distribution built of delta-functions
with imaginary support is obvious nonsense.

We can discover the sense hiding behind this nonsense if we remember the
dependence on $\hbar$ of the semiclassical approximation.  In leading
semiclassical approximation, the probability distribution for the electric
field at a given point (or indeed for any dynamical variable) is of the form
$$\rho(E)= \sum_i A_i(E)e^{B_i(E)/\hbar}~,
\eqn\fieldG
$$
where the $B's$ are independent of $\hbar$ and the $A$'s are monomials in
$\hbar$. In our case, there are three terms in this series.  One is the
leading one; the others are exponentially suppressed, but are still retained
because they are the leading charge-dependent terms.  If one of the $B$'s is a
real function with a maximum on the real axis, the corresponding term in
$\rho$ simulates a delta-function in the small-$\hbar$ limit.  However, even
if $B$ is a complex function with a complex stationary point, this still
simulates a delta-function (with {\it complex} support) provided the integral
of interest has sufficient analyticity that the contour of integration
can be distorted through the stationary point.  An example is a Gaussian with
complex center, $B_i=-{1\over 2}(E-z)^2$. The contribution of this term to
$\langle E^n \rangle$ is proportional to $z^n$, for arbitrary complex $z$.

Of course, we don't know the detailed shape of $B(E)$ from our computations.
All we know is $E_0$ and $B(E_0)$.  (Even if we had computed $C(\beta\hbar)$
all it would have told us is the value of $A(E_0)|B''(E_0)|^{-1/2}$.)
Nevertheless, the general situation is clear. The charge-dependent term in
$\rho(E)$ acts in the small-$\hbar$ limit like a sum of delta-functions with
imaginary support, but in fact, it is nothing of the kind; it is something
like a Gaussian with complex center, a very rapidly oscillating function
with very many very closely spaced maxima and minima all along the real axis.

\section{The Vortex and the No-Hair Theorem}
In the context of the abelian Higgs model that we have been
analyzing in
this section, the no--hair theorem states the
following:\Ref\adler{S.
L. Adler and R. B. Pearson, Phys. Rev. {\bf D18} (1978)
2798.}  If
$v^2>0$,
then any
stationary (Lorentzian) black hole solution, such that all
gauge--invariant observables are non-singular
both at the horizon and at spatial infinity, must
have a  vanishing electromagnetic
field, and a covariantly constant Higgs field,
outside the event horizon.

The vortex solution that we have exhibited here demonstrates
that there
is no corresponding statement about {\it Euclidean} black
holes.
The vortex is stationary in the sense that gauge--invariant
quantities
are $\tau$-independent, and it is analytic on the minimal
``horizon''
two sphere at the origin of the $r$-$\tau$ plane.  Thus,
there {\it do}
exist non-singular stationary solutions to the imaginary
time field
equations that have topology $R^2\times S^2$ and are
asymptotically flat,
with a nontrivial $F_{r\tau}$ and an $r$-dependent
$|\phi|^2$.

The existence of the Euclidean vortex solution is consistent
with the
no--hair theorem, because when the vortex is continued back
to real
time, it fails to satisfy the hypotheses of the no--hair
theorem.  For
example, the electric field is imaginary.

We also note that, in a certain sense, the Euclidean vortex
is {\it not}
static.  Although gauge--invariant local observables are
independent of
$\tau$, there is no nonsingular gauge in which the Higgs
field $\phi$ is
$\tau$--independent, since the phase of the Higgs field
advances by
$-2\pi k$
as $\tau$ increases by $\beta\hbar$.  The vorticity $k$
is a global property of the solution that changes sign under
the
transformation $\tau\to-\tau$.  Correspondingly, in any
nonsingular
gauge, if we perform the naive continuation to real time,
the gauge--invariant quantity $|\phi|^2$ becomes time
dependent.

Expectation values of observables on the black hole
background are obtained by summing contributions from the
various vorticity sectors, as described in Section~4.5.
These expectation values are static, and have suitable
reality properties, when continued to real time.  In accord
with the no--hair theorem, though, the expectation values do
not solve the classical field equations; since the field
equations are nonlinear, a sum of solutions is not a
solution.  In short, by performing the sum over $k$, we
proceed from non-static solutions to static non-solutions,
thus violating the spirit of the no--hair theorem while
respecting its mathematical content.

We should also remark that the status of the no--hair
theorem for the abelian Higgs system remains unsettled, as
was recently stressed by Gibbons.\Ref\giblisbon{G. W.
Gibbons,
``Self-Gravitating Magnetic Monopoles, Global Monopoles
and Black Holes,'' Lectures at the 1990 Lisbon Autumn
School,
DAMTP R90/31 (1990).}  The (classical) no--hair
property seems very plausible, but no fully satisfying and
sufficiently general proof has yet been found.

We conclude this discussion with some further remarks
concerning the $\mu\to 0$ limit.  For classical {\it
Lorentzian} black holes, this limit is highly singular.  If
$\mu=0$, then a stationary black hole can have an electric
field.  But for any nonzero $\mu$---however small---no
electric field is allowed.

One might hope that black hole physics will behave more
smoothly in the $\mu\to 0$ limit, when the effects of
quantum hair are properly taken into account.  Indeed, for
{\it Euclidean} solutions, as we have seen, the situation
{\it is} different.  Our vortex solution smoothly approaches
the corresponding solution (with the same value of
$\omega=2\pi k/N$) for a black hole coupled to  massless
electrodynamics.

Nevertheless, the discontinuous behavior of black hole
physics in the $\mu\to 0$ limit persists.  We have found
that the expectation value of the electric field outside a
black hole that carries $Z_N$ charge, while nonvanishing, is
far from classical.  It is, in fact, exponentially small for
small $\hbar$.  In contrast, if $\mu=0$, the expectation
value is $\hbar$-independent.

This discontinuous behavior seems less mysterious, though,
when we recognize that the semiclassical limit has been
taken in a much different way for a black hole with $Z_N$
charge than for a black hole with ordinary electric charge,
as we emphasized in Section~4.2.  A black hole with
specified $Z_N$ charge has fixed $Q/N\hbar e$, and so the
electric charge $Q$ is of order $\hbar$.  We can hardly
expect this object to behave like a black hole with a
specified classical charge, in the limit $\hbar\to 0$.

If we want the semiclassical physics of a black hole with
quantum hair to behave smoothly  in the $\mu\to 0$ limit,
then, we must define the semiclassical limit differently
than before.  In particular, we should allow N to become
large; then the sum in eq.~\NprojB\ approximates the
integral in eq.~\projM.  However, there is a catch.  To
justify replacing the sum in eq.~\NprojB\ by an integral as
$N\to \infty$, we require
$$
Z(\beta,k+1)-Z(\beta,k)\to 0~.
\eqn\hairA
$$
But, for $v^2\to 0$ (and neglecting gravitational back
reaction), we have
$$
Z(\beta,k)\simeq \exp\left(-\Delta S_k^{(thick)}\right)
\simeq \exp\left(-~{\pi k^2\over 2\hbar(Ne)^2}\right)~.
\eqn\hairB
$$
So $Z(\beta,Q)$ in eq.~\NprojB\ will not be well
approximated by the corresponding expression in massless
electrodynamics unless
$$
N\hbar e>>Q~,~~~~~\hbar (Ne)^2>>1~.
\eqn\hairC
$$
The catch is that $\hbar(Ne)^2$ is a loop expansion
parameter.  Thus, if eq.~\hairC\ is satisfied, the Higgs
model is strongly coupled, and the semiclassical evaluation
of $Z(\beta,k)$ cannot be justified.

The conclusion is that, while it is conceivable that,
in a suitable limit, the
physics of a black hole with screened electric hair can
match
smoothly with the physics of a black hole with unscreened
hair, this cannot happen within the
domain of validity of the semiclassical approximation.

In the next section, we will find that, in the case of {\it
magnetic} quantum hair, the situation is quite different.

\chapter{Screened Magnetic Charge}

\section{$Z_N$ Magnetic Charge Projection}
As we have already noted in Sections 2 and 3, a gauge theory
with an unbroken $SU(N)/Z_N$ gauge symmetry can contain
magnetic monopoles with $Z_N$ magnetic charges.  In the
classical approximation, such a monopole has an infinite
range magnetic field.  But if $SU(N)/Z_N$ is not
spontaneously broken, we expect that, due to quantum
effects, the vacuum of the theory is magnetically
disordered.  Hence the theory exhibits color
confinement---electric
fields are confined to stable flux tubes, and
magnetic fields are screened.  The magnetic field of a
monopole decays like $e^{-\mu r}$ at long range, where
$\hbar\mu$ is the glueball mass gap.

Nevertheless, in spite of the magnetic screening, the $Z_N$
magnetic charge of an object can be detected in principle at
arbitrarily long range, by means of the nontrivial
Aharonov--Bohm interaction of a magnetic charge with an
electric flux tube.  It follows that, despite the screening,
a stationary black hole can carry magnetic
charge.\refmark{\prekra}  Our
objective in this section is to perform a semiclassical
analysis of the effect of $Z_N$ magnetic charge on the
physics of a black hole.\refmark{\cpwA,\cpwB}

Evidently, the detection of $Z_N$ magnetic charge with an
electric flux tube is strikingly similar to the detection of
$Z_N$ electric charge with a cosmic string.  Since we have
already analyzed the effects of $Z_N$ electric charge on a
black hole in much detail, one may wonder whether it is
really necessary to repeat the whole analysis for the case
of magnetic charge.  In fact, we will find that the
semiclassical theory of a black hole with $Z_N$ magnetic
charge is considerably different than the theory that we
have already developed for a black hole with $Z_N$ electric
charge.  The origin of the difference is already noted in
the preceding discussion:  While the screening of electric
fields due to the Higgs mechanism is a classical phenomenon
(zeroth order in $\hbar$), the screening of magnetic field
due to confinement is a quantum effect (actually
nonperturbative in $\hbar$).

To calculate the effect of $Z_N$ magnetic charge on black
hole thermodynamics, we must first devise a Euclidean path
integral prescription for computing the partition function
restricted to a particular magnetic charge sector.  For the
case of $U(1)$ magnetic charge discussed in Section~3.7, we
saw that a magnetic charge projection was
implemented trivially---the sector with magnetic charge
$P$ is obtained
by restricting the path integral to configurations with
magnetic flux $4\pi P$ at $r=\infty$.  In spite of the
screening, a similar procedure applies in the case of $Z_N$
magnetic charge.  As explained in Section~2.4, the total
magnetic charge can be characterized by the topology of the
gauge field (or, alternatively, the Higgs field) on the
two--sphere at $r=\infty$.  The charge projection is
performed by restricting the field configurations to the
appropriate topological class.

\section{Electric--Magnetic Duality}
Because of the strong similarity between $Z_N$ electric and
$Z_N$ magnetic charge, it seems odd that the charge
projections in the two cases are so different.  We saw that
the expression eq.~\NprojB\ for $Z(\beta,Q)$ has a very
appealing interpretation.  We may think of the vorticity $k$
as the number of times that a virtual cosmic string world
sheet wraps around the horizon of the black hole, and the
phase multiplying $Z(\beta,k)$ is just the Aharonov--Bohm
phase acquired by this virtual string.  Surely, we should be
able to understand the dependence of black hole
thermodynamics on the $Z_N$ magnetic charge in a similar
way, for the effect of the magnetic charge is to weight the
contributions due to virtual electric flux tubes by
appropriate Aharonov--Bohm phases.  One suspects, therefore,
that there is an alternative way to construct the $Z_N$
magnetic charge projection that is more closely parallel to
what we did for $Z_N$ electric charge.

This is indeed the case, as we can see by performing an
electric--magnetic duality transformation.\Ref\Hooftdual{G.
't Hooft, Nucl. Phys. {\bf B153} (1979) 141.}  In an $SU(N)$
gauge theory with an unbroken local $Z_N$ symmetry, states
can be classified according to how they transform under
$Z_N$.  The $Z_N$ electric charge $\tilde e$ is defined by
$$
U(\tilde m)\ket{\tilde e}=e^{2\pi i \tilde m\tilde
e/N}\ket{\tilde e}~,
\eqn\magprojA
$$
where $U(\tilde m)$ represents the $Z_N$ transformation
$e^{2\pi i \tilde m/N}$; this charge takes the values
$$\tilde e=0,1,2,\dots,N-1~.
\eqn\magprojB
$$
We may insert the projection operator
$$
P(\tilde e)={1\over N}\sum_{\tilde m=0}^{N-1}e^{-2\pi i
\tilde m
\tilde e/N}U(\tilde m)
\eqn\magprojC
$$
into the path integral, as in Sections 3.3 and 4.1, to
obtain an expression for the partition function in the
sector with charge $\tilde e$;  the result is
$$
Z(\beta,\tilde e)={1\over N}\sum_{\tilde m=0}^{N-1}e^{-2\pi
i \tilde m
\tilde e/N} Z(\beta, \tilde m)~,
\eqn\magprojD
$$
where $Z(\beta, \tilde m)$ is a path integral over
configurations that satisfy the constraint
$$
\left[P\exp\left(ie\int_0^{\beta\hbar}d\tau~A_\tau(\tau,\vec
x)
\right)
\right]_{r=\infty}=e^{2\pi i \tilde m /N}~.
\eqn\magprojE
$$
In the black hole sector, with topology $R^2\times S^2$, we
may think of $\tilde m$ as the topological magnetic flux, or
{\it vorticity}, in the $r$--$\tau$ plane.
Naturally, we can invert the Fourier transform and write
$$
Z(\beta,\tilde m)=\sum_{\tilde e=0}^{N-1}e^{2\pi i \tilde m
\tilde e/N}Z(\beta,\tilde e)~.
\eqn\magprojF
$$

Now consider the case of an $SU(N)/Z_N$ gauge theory that
contains no dynamical $Z_N$ electric or magnetic charges.
Then we may think of $\tilde e$ as the electric flux on each
$S^2$ parametrized by $\theta$ and $\phi$.  And imagine that
the $r$--$\tau$ plane is compactified to a two--sphere, so
that the Euclidean black hole geometry has the topology
$S^2\times S^2$.  The way to derive a duality relation is to
interchange the two two--spheres,
$$
S^2(\theta,\phi)~\longleftrightarrow ~S^2(r,\tau)~.
\eqn\magprojG
$$
In other words, we may reinterpret eq.~\magprojF, regarding
$\tilde m$ as the magnetic flux on $S^2(\theta, \phi)$, and
$\tilde e$ as the electric flux on $S^2(r,\tau)$.  So, at
least in the case where there are no dynamical magnetic
monopoles, the partition function of a black hole with
specified $Z_N$ magnetic charge $\tilde m$ is obtained by
summing  the sectors with various values of the {\it
electric} vorticity $\tilde e$, weighted by appropriate
phases.  The electric vorticity  is the number of times that
a virtual electric flux tube wraps around the black hole
horizon, and the associated phase is just the Aharonov--Bohm
phase acquired by the virtual flux tube.  This is the result
we expected.

\section{The Thin String Limit}
In a confining gauge theory, the thickness $\mu^{-1}$ and
tension $T_{\rm string}$ of an electric flux tube are, in
order of magnitude,
$$
\mu^{-1}\sim {\hbar\over\Lambda}~,~~~~
T_{\rm string}\sim {\Lambda^2\over \hbar}~,
\eqn\thinmagA
$$
where $\Lambda$ is the characteristic mass scale of the
theory.  Of course, the tension and mass gap are really
nonperturbative in $\hbar$; we may express the mass scale
$\Lambda$ as
$$
{\Lambda\over\hbar}\sim a^{-1}~\exp\left[
-~{1\over b~\hbar e^2(a)}\right]~,
\eqn\thinmagB
$$
where $a$ is a short--distance cutoff, and $e^2(a)$ is the
bare gauge coupling.  There is no confinement, and no
magnetic screening (at zero temperature), to any finite
order in the $\hbar$ expansion.

Now consider the effect of $Z_N$ magnetic charge on a black
hole, in the case where the screening length $\mu^{-1}$ is
very small compared to the size $r_+$ of the black hole (the
``thin string limit'').  The classical black hole solution
has a long--range magnetic field, but the classical
approximation is badly misleading in the thin string limit;
nonperturbative effects screen the magnetic field, and
nearly extinguish it entirely.

The leading charge--dependent effects in the thin string
limit are due to a virtual electric flux tube that wraps
once around the event horizon, with either orientation.
These effects are most easily analyzed, then, using the dual
formulation described in the previous section; the $\tilde
e=\pm 1$ sectors dominate the charge--dependence in
eq.~\magprojF.  Of course, since nonperturbative effects are
important, the evaluation of the charge dependence of the
free energy is not, strictly speaking, semiclassical.  But
if we imagine working with an effective field theory that
describes the infrared behavior of the confining theory,
then the analysis becomes virtually identical to that
described in Section~4.3.

For $\mu r_+>>1$, the path integral in the sectors with
$\tilde e=\pm 1$
is dominated by a configuration with a pointlike electric
vortex sitting at the origin of the $r$-$\tau$ plane.  In
this
configuration, the effective matter field action is
$$
S_{\rm matter}^{(eff)}\simeq 4 \pi r_+^2 T_{\rm string}~,
\eqn\thinmagC
$$
as in eq.~\thinA.  Proceeding as in Section~4.3, we find the
leading
charge--dependent contribution to the black hole
temperature,
$$
\beta^{-1}(M,\tilde m)-\beta^{-1}(M,\tilde m)\propto
\left[1-\cos\left({2\pi \tilde m\over N}\right)\right]~
\exp\left(-16\pi(GM)^2 T_{\rm string}/\hbar\right)~
\eqn\thinmagD
$$
(for $GT_{\rm string}<<1$).

The virtual electric flux tubes that sweep around the black
hole generate a radial magnetic field outside the horizon,
as in the discussion in Section~4.5.  Of course,
$F_{\theta,\phi}$ is not gauge invariant in a nonabelian
theory, but we can characterize the field by considering,
for example, the expectation value of the gauge--invariant
operator ${\rm tr}(F_{\theta,\phi}^2)$.  This behaves as
$$
\eqalign{
&\VEV{{\rm tr}~F_{\theta,\phi}(r)^2}_{M,\tilde m}\sim
{\rm charge~independent}\cr
&~~~~+~\cos\left({2\pi\tilde m\over N}\right)
{}~\exp\left(-16\pi(GM)^2 T_{\rm string}/\hbar\right)~
e^{-2\mu r}~.\cr}
\eqn\thinmagE
$$
The charge--dependent field is suppressed not just by the
screening factor $e^{-2\mu r}$, but also by the tunneling
factor $\exp(-\Delta S_{\rm vortex}^{(eff)}/\hbar)$.  Notice
that its dependence on $\tilde m$ is different than that of
the classical solution described in Section~3.8; it is
proportional to $\cos(2\pi \tilde m/N)$ instead of $\tilde
m(N-\tilde m)$.

\section{The Thick String Limit}
Now we consider the opposite limit, in which the thickness
$\mu^{-1}$ of an electric flux tube is much larger than the
size $r_+$ of the black hole.  In this limit, it is more
convenient to implement the magnetic charge projection by
restricting the path integral to configurations with
specified $Z_N$ flux $\tilde m$.  Then the evaluation of the
black hole partition function, in the leading semiclassical
approximation, proceeds much as in the magnetic
Reissner--Nordstr\"om case discussed in Section~3.7.  The path
integral is dominated by the magnetically charged black hole
classical solution that we described in Section~3.8 (or, if that
solution is unstable, by a solution analogous to the one constructed in
Ref.~\Nair).

Of course, we are obligated to consider the nonperturbative
corrections.  In the thin string limit, these completely
invalidate the leading semiclassical result, as we saw
above.  But in the thick string limit, the nonperturbative
corrections to the semiclassical calculation are small.  The
Yang--Mills action of the magnetically charged black hole is
dominated by the magnetic field close to the event horizon.
In the thick string limit, asymptotic freedom ensures that
the  gauge coupling renormalized at the distance scale $r_+$
is small; hence, quantum fluctuations about the classical
solution are suppressed near the horizon.  Nonperturbative
quantum corrections {\it are} important at a distance of
order $\mu^{-1}$ from the black hole, where the magnetic
field begins to become screened.  But the nonperturbative
contribution to the Yang--Mills effective action is
suppressed, relative to the classical contribution, by a
factor of order $(\mu r_+)^2$, and so can be neglected in
the thick string limit.

Therefore, in the thick--string limit, it is a good approximation to
saturate the path integral with the appropriate Euclidean classical
solution.  For values of the parameters such that the
the solution described in Section 3.8 is stable, then,
the thermodynamics of a black hole with $Z_N$
magnetic charge $\tilde m$, in the thick string limit and
the leading semiclassical approximation, is identical to
that of a Reissner--Nordstr\"om black hole, where the
effective charge $P_{\tilde m}$ of the hole is
$$
P_{\tilde m}^2=\left({1\over 2 e^2}\right)
{\tilde m(N-\tilde m)\over N}
\eqn\thinmagF
$$
(as in eq.~\ymD; $e^2$ denotes the gauge coupling
renormalized at distance scale $r_+$).  The effect of the
charge on the thermodynamics can be quite significant;  it
can even lower the temperature to zero and shut down the
Hawking evaporation of the hole.  Furthermore, outside the
event horizon, and for $r<<\mu^{-1}$, the black hole has a
magnetic field that is well approximated by that of the
classical solution.

\section{Screened Magnetic Charge vs. Screened Electric
Charge}

As we have seen, the physical effects of screened electric
and magnetic charge on a black hole are similar in the thin
string limit, but dramatically different in the thick string
limit.  Why?

The difference arises because of the different role played
by $\hbar$ in the two cases.  The electric charge quantum is
of order $\hbar e$.  So $Z_N$ electric charge is really
quantum--mechanical, and becomes irrelevant in the classical
limit.  But the magnetic charge quantum is of order $1/e$.
So $Z_N$ magnetic charge can have effects that survive in
the classical limit.

Furthermore, since the magnetic flux carried by a cosmic
string is of order $1/e$,  the cosmic string is a classical
object, and virtual strings arise as quantum--mechanical
fluctuations only very rarely.  In contrast, an electric
flux tube carries a flux of order $\hbar e$, and virtual
strings occur copiously as quantum fluctuations.
The dependence of black hole physics on $Z_N$ electric
charge is dominated
by configurations such that a single string world sheet
wraps around the black hole, even if the string is very
thick.   But, because the electric flux tube carries a small
flux, the cost in effective action of adding another flux
tube, in the thick string limit, is correspondingly small.
Thus, the dependence on the $Z_N$ magnetic charge is
dominated by configurations that contain not a single
string, but many.  We may think of the classical magnetic
field  as resulting from the ``condensation'' of the cloud
of virtual strings that surrounds the black hole.

Since an electric flux tube is really a nonperturbative
object,
it may seem miraculous that the effects of virtual electric
flux tubes admit a classical description.  Asymptotic
freedom makes this miracle possible.  The effective action
of the cloud of virtual strings surrounding the hole is
dominated by the fields close to the event horizon, where
they may be regarded as weakly coupled.

The sharp distinction between screened electric fields and
screened magnetic fields applies if we insist on analyzing
both in the semiclassical approximation.  A confining theory
that is weakly coupled at the event horizon of a black hole
exhibits much different physics than a Higgs theory that is
weakly coupled at the event horizon.  But the distinction
blurs as the confinement length scale becomes comparable to
the size of the hole, so that virtual electric flux tubes
that envelop the hole become suppressed, or if the Higgs
theory is strongly coupled, so that virtual cosmic strings
are unsuppressed.

Our analysis of screened magnetic charge enables us to
reexamine an issue that we previously addressed in
Section~4.6:  Is it possible for the physics of a black hole
with screened hair to match up smoothly, as the inverse
screening length $\mu$ approaches zero, with the physics of
a black hole with unscreened hair?  In the case of $Z_N$
electric hair, we saw that this is not possible if the Higgs
theory is weakly coupled.  Insofar as  a strongly--coupled
abelian
Higgs theory resembles a weakly--coupled nonabelian theory,
it seems appropriate to reopen the question now.

Suppose that we introduce into an $SU(N)/Z_N$ gauge theory
the necessary Higgs structure so that, by appropriately
adjusting the parameters of the Higgs potential, we may
break the gauge group to the abelian subgroup $U(1)^{N-1}$.
In this ``Coulomb phase'' of the model, there are black hole
solutions that carry $U(1)^{N-1}$ charges.   Among these
solutions are ones that have the same magnetic field as the
black hole solutions of a model with unbroken $SU(N)/Z_N$,
the solutions that  we described in Section~3.8.  But, in
the Coulomb phase there is no confinement and  no magnetic
screening; the classical magnetic field survives at
arbitrarily long range.

Now, in the semiclassical approximation, the black hole
partition function in a sector with specified magnetic
charge is dominated by the corresponding
magnetically--charged solution.  This is true
in both the Coulomb phase
and the confining phase of the model. Furthermore, the black
hole solutions in the Coulomb phase have a covariantly
constant Higgs field.  (The instability described in Ref. \Nair\ does
not apply here.  The Coulomb field extends all the way into the horizon
whether or not the Higgs field that breaks $SU(N)/Z_N$ to $U(1)^{N-1}$
turns on outside the horizon.)
Thus, the leading semiclassical thermodynamics of a
magnetically--charged black hole in the confining phase is
identical to the leading semiclassical thermodynamics of a
charged black hole in the Coulomb phase.

If the Higgs symmetry breaking scale $v$ is small, so that
effects
that depend on the details of the Higgs potential may be
neglected, the semiclassical expansion of the partition
function is the same, to each order in $\hbar$, in the
Coulomb phase as in the confining phase.  Finally, we recall
that nonperturbative corrections in the confining phase are
small when the confinement length scale is much larger
than the black hole.  We conclude that the physics of a
magnetically charged black hole in the confining phase, in
the limit $\mu r_+<<1$, coincides with the physics of a
magnetically charged black hole in the Coulomb phase, in the
limit $ev r_+<<1$.  In this instance, no annoying
discontinuity is encountered as the inverse screening length
$\mu$ goes to zero.

Our study of screened electric and magnetic hair on black
holes may be summarized as follows:  The no--hair theorems
assert that {\it classical} hair is incompatible with {\it
classical} screening.  But they leave open two possible ways
to evade the dictum that a black hole has no (screened)
hair---either the hair or the screening (or, conceivably,
both) may be quantum mechanical.  In a weakly--coupled Higgs
system, the screening of electric fields is classical in the
sense that the screening length $\mu^{-1}$ is independent of
$\hbar$.  We have seen that this classical screening can be
reconciled with a non-vanishing electric field outside the
event horizon of a black hole, provided that the field
disappears in the $\hbar\to 0$ limit.  On the other hand, in
a confining gauge theory, the screening of magnetic fields
is quantum--mechanical in the sense that $\mu\to 0$ as
$\hbar\to 0$.  We have seen that this non-classical
screening is compatible with a classical
($\hbar$-independent) magnetic field outside the horizon.

\chapter{Dual Formulation of Broken Symmetry Phases}

In this section we shall discuss the representation of
Goldstone boson,
St\"uckelberg-Higgs, and axion theories in dual form.  The
use of
such alternative representations can be a source of
insight
(or confusion) in a wide variety of problems, including but
not restricted
to black hole physics.  We will now work in Lorentzian
spacetime, unless otherwise noted.

\section{Goldstone boson}
To begin consider the theory of a massless scalar
(Goldstone) field
$\phi$, described by the Lagrangian
$$
{\cal L} = {F^2\over 2} (\partial_\mu \phi )^2 .
\eqn\dualaa
$$
(For simplicity we shall write our formulas
in the form appropriate for flat space. Their generalization
to
curved space is more or less immediate; see below.)
This Lagrangian may also be obtained from
$$
{\cal L} = i\partial_\mu \phi\, J^\mu + {1\over 2F^2}
J_\mu^2
\eqn\dualba
$$
by eliminating $J_\mu$, which appears only algebraically.
Indeed by setting the variation with respect to $J$ equal to
zero we find
$$
J_\mu = -i F^2 \partial_\mu \phi ,
\eqn\dualca
$$
and substituting this into \dualba\ yields \dualaa .

The same equations of motion
result if we add a total derivative to
the action.  Thus
we will obtain the same equations of motion if we integrate
the first term on the right-hand side of \dualba\ by parts,
and
drop the surface term.  This procedure gives us the
Lagrangian
$$
{\cal L} = -i\phi\, \partial_\mu J^\mu + {1\over 2F^2}
J_\mu^2
\eqn\dualda
$$
in which $\phi$ appears only algebraically.  The variation
with respect
to $\phi$ simply gives us the constraint
$$
\partial_\mu J^\mu = 0
\eqn\dualea
$$
expressing conservation of the current $J$.  The Lagrangian
is
just
$$
{\cal L} = {1\over 2F^2} J_\mu^2 ,
\eqn\dualfa
$$
subject, of course, to the constraint \dualea .

In three-plus-one space-time dimensions one may solve the
constraint
by writing
$$
J^\mu = {-i\over 2} \epsilon^{\mu\nu\rho\sigma} \partial_\nu
B_{\rho\sigma}
\eqn\dualga
$$
where $B$ is an antisymmetric tensor field.
Substituting this into \dualfa , we see that the Goldstone
boson theory
has been re-written in terms of a two index antisymmetric
tensor field.

In curved space one starts with
the Goldstone boson Lagrangian
$$
{\cal L} =
{F^2\over 2} \sqrt{g} g^{\mu\nu} \partial_\mu \phi
\partial_\nu \phi .
\eqn\dualha
$$
One may obtain this from
$$
{\cal L} = i \partial_\mu \phi J^\mu
+ {1\over 2F^2 \sqrt{g}} g_{\mu\nu} J^\mu J^\nu .
\eqn\dualgaa
$$
In this formulation $J^\mu = \sqrt{g} g^{\mu\nu}
\partial_\nu \phi$
is a vector {\it density}
which satisfies $\partial_\mu J^\mu = 0$ identically.
$B$ may be introduced as before, to solve the constraint; it
is
a proper two-index tensor (the numerical epsilon symbol is a
tensor density,
and absorbs the excess $\sqrt{g}$).

Let us now briefly consider how this formalism
applies to the theory of a complex scalar
field.  Let $\eta$ be a complex scalar described by the
Lagrangian
$$
{\cal L} = |\partial_\mu \eta|^2 - V(|\eta |)
\eqn\dualab
$$
where $V$ is the potential term.  Writing $\eta$ in
polar form $\eta = \rho e^{i\phi}$ we have
$$
{\cal L} =
(\partial_\mu \rho)^2  + \rho^2 (\partial_\mu \phi)^2 -
V(\rho )\, .
\eqn\dualbb
$$
Now if $\rho$ develops an expectation value of magnitude
$F/\sqrt 2$, we
simply expand $\rho \equiv {1\over \sqrt 2} (F + \zeta)$.
Whereas $\rho$ was constrainted to be $\geq 0$, we have
$-F \leq \zeta$.
The $\phi$-dependent terms
in \dualbb\ are then simply
$$
{\cal L} =
{1\over 2} (F + \zeta)^2 (\partial_\mu \phi)^2 \, .
\eqn\dualcb
$$
Now we can introduce the dual description much as before.
The Lagrangian will still be of the current-squared form,
but the constraint for the dual current will
be altered, to assume the form
$$
\partial_\mu \bigl(J^\mu (1+ {\zeta\over F}) \bigr) = 0\, .
\eqn\dualdb
$$
This can be solved in the form
$$
J^\mu = {-i\over 2} {\epsilon^{\mu\nu\rho\sigma}
\partial_\nu B_{\rho\sigma}
         \over (1 + \zeta /F) } \,  .
\eqn\dualeb
$$
This expression becomes singular where $\zeta$ approaches
its
minimum (\ie\ where the magnitude $\rho$ of the original
scalar field
vanishes).

Thus the dual formulation can be extended to the full
complex scalar
theory, in a rather straightforward fashion.
Also, the larger theory
will tell us how to proceed, in principle,
at the singular points where
$\rho = {1\over \sqrt 2} (F + \zeta )$
vanishes -- such as occurs at the
center of cosmic strings, where the earlier formulation
breaks
down.

\section{Gauge coupling}
The Goldstone boson Lagrangian \dualaa\
is invariant under the transformation
$\phi \rightarrow \phi + \lambda$, where $\lambda$ is a
numerical
constant.  We can promote this to a local symmetry
transformation by
coupling to a gauge field with appropriate transformation
properties.  Indeed
$$
{\cal L} = {F^2\over 2} (\partial_\mu \phi + eA_\mu )^2
\eqn\dualna
$$
is invariant under the local transformation
$$
\eqalign{
\phi &\rightarrow \phi + e\lambda \cr
A_\mu &\rightarrow A_\mu - \partial_\mu \lambda \cr}
\eqn\dualoa
$$
where $\lambda$ is a function on space and time.

The gauge freedom can be used to set $\phi = 0$.  Then
\dualna\ reduces to ${1\over 2}(eF)^2 A_\mu^2$, that is to a
mass term
for $A$.  This is the Higgs mechanism in its simplest form,
as
first presented by St\"uckelberg.  It is also essentially
equivalent
to the London theory of superconducting electrodynamics.
As we have just seen in the global symmetry case,
$\phi$ can be regarded as the phase of an ordinary complex
scalar
field, that acquires a vacuum expectation value.  This
explains the
additive form of the gauge transformation law \dualoa .
(The
St\"uckelberg-London Lagrangian is an approximation to the
usual
Landau-Ginzburg Lagrangian, in which the magnitude of the
scalar field is
frozen and only its phase allowed to vary.)

To construct the dual form of this theory, we introduce
$$
{\cal L} = iJ^\mu\, (\partial_\mu\phi + eA_\mu) +
{1\over 2F^2} J_\mu^2 .
\eqn\dualpa
$$
It is easy to verify that elimination of $J$ from
\dualpa\ leads right back
to \dualna .  Alternatively one may integrate by parts, and
eliminate
$\phi$.  This leads to the constraint $\partial_\mu\, J^\mu
= 0$ and
to the simple Lagrangian
$$
{\cal L} = ieJ^\mu\, A_\mu + {1\over 2F^2}J_\mu^2
\eqn\dualqa
$$
with $J_\mu$ subject to the conservation constraint.

As before, we may solve the constraint by writing $J$ in the
form
\dualga\ .  This leads to the Lagrangian
$$
{\cal L} = {e\over 2} \epsilon^{\mu\nu\rho\sigma}\, A_\mu
\partial_\nu B_{\rho\sigma}
- {1\over 2F^2} (\epsilon^{\mu\nu\rho\sigma}\, \partial_\nu
B_{\rho\sigma})^2.
\eqn\dualra
$$
Thus the difference between
the Goldstone and the Higgs theories, in this dual
form, is simply the absence or presence of the first term on
the right-hand
side of \dualra\ .  Indeed this term is just the
current-gauge field coupling,
written in a slightly unfamiliar form.  It may also be
written in another,
suggestive
way after an integration by parts:
$$
{e\over 2}\epsilon^{\mu\nu\rho\sigma}\,
A_\mu \partial_\nu B_{\rho\sigma}
\rightarrow {e\over 4}\epsilon^{\mu\nu\rho\sigma}\,
F_{\mu\nu} B_{\rho\sigma}.
\eqn\dualsa
$$
In the first form the potential $A$ of the gauge field
occurs, but
only the combination $H_{\mu\nu\rho} = \partial_\mu
B_{\nu\rho} +
{\rm cyclic~ permutations}$ -- \ie\ , the field strength --
for the
antisymmetric tensor appears.  In the second form
the roles are reversed.  Only the field strength for the
gauge field occurs,
but the bare potential $B$ appears.
In the second form, the simple charge coupling
has come closely to resemble the $\theta$ term of gauge
theories.
This form of the gauge coupling in the Higgs theory will
play a central role
in our analysis in the following
chapter.

In the next chapter we shall need the curved-space form of
the
dualized gauge theory.  For future reference, then, let us
record
the generalizations of \dualpa\ , \dualra\ :
$$
{\cal L} ~=~
  iJ^\mu (\partial_\mu \phi + eA_\mu) ~+~
  {1\over 2F^2 \sqrt g}g_{\mu \nu} J^\mu J^\nu
\eqn\dualsb
$$
$$
{\cal L} ~=~ {e\over 4}\epsilon^{\mu \nu \rho \sigma}
F_{\mu\nu} B_{\rho \sigma}
       ~+~ {1\over 2F^2 \sqrt g} g_{\mu \nu} J^\mu J^\nu~.
\eqn\dualsc
$$
The contribution of the first term in \dualsc\ to the action
is actually independent of the spacetime metric; it can be
expressed in terms of a wedge product of differential forms.
The term $S_{B\wedge F}$ that couples the gauge field to the
antisymmetric tensor field has the form
$$
e^{iS_{B\wedge F}/\hbar}=\exp\left( {ie\over \hbar}
\int B\wedge F\right)~.
\eqn\Scoup
$$

\section{Axion}
As our final example we consider the dual formulation of
axion physics.
The basic Lagrangian (for an axion coupled to an abelian
gauge field) is
$$
{\cal L} = {f^2\over 2}(\partial_\mu \phi)^2
+ \phi\epsilon^{\mu\nu\rho\sigma} F_{\mu\nu}F_{\rho\sigma} .
\eqn\dualta
$$
In the now familiar way, we obtain this from the alternative
Lagrangian
$$
{\cal L} = iJ^\mu\, \partial_\mu \phi + {1\over 2f^2}J_\mu^2
+
\phi\epsilon^{\mu\nu\rho\sigma} F_{\mu\nu}F_{\rho\sigma}
\eqn\dualua
$$
by eliminating $J$.  Alternatively, integrating by parts and
eliminating $\phi$ yields the constraint
$$
\partial_\mu J^\mu = i \epsilon^{\mu\nu\rho\sigma}
F_{\mu\nu}F_{\rho\sigma}.
\eqn\dualva
$$
This constraint may be solved in the form
$$
J^\mu = {-i\over 2} \epsilon^{\mu\nu\rho\sigma} \partial_\nu
B_{\rho\sigma}
+ {i\over 2} \epsilon^{\mu\nu\rho\sigma} A_\nu
F_{\rho\sigma}.
\eqn\dualwa
$$
In terms of this constrained $J$, the Lagrangian is simply
${\cal L} = {1\over 2f^2} J_\mu^2$.  Upon writing out this
Lagrangian
in terms
of the antisymmetric tensor $B$, one finds peculiar
cross-couplings
between $B$ and the gauge fields, which
are not manifestly gauge invariant.  They are closely
related to
the Green-Schwarz terms that play an important role in
superstring
theory.\Ref\schwarz{M. B. Green and J. H. Schwarz, Phys.
Lett.
{\bf 149B} (1984) 117.}  From the point of view of axions, they
reflect
fact that exchange of the axion between photons involves
a $1\over q^2$ propagator which partially cancels the
derivatives
coming from the $\phi\epsilon^{\mu\nu\rho\sigma}
F_{\mu\nu}F_{\rho\sigma}$
vertices.

The gauge transformation property of $B$ is peculiar.  It
follows from the gauge invariance of $J$ in \dualwa .  Thus
under
a gauge transformation
we have
$$
\eqalign{
B_{\mu\nu} &\rightarrow B_{\mu\nu} - \lambda F_{\mu\nu} \cr
A_\mu &\rightarrow A_\mu - \partial_\mu \lambda . \cr}
\eqn\dualxa
$$

\chapter{Duality and Hair}

In this chapter we shall look at quantum hair from a
different perspective, using the dual formalism developed
in the previous chapter.  We shall find, of course,
that this formalism yields a superficially
different but fully equivalent description of the earlier
results.  The black hole hair will be associated with the
{\it fractional part\/} of the surface integral of the
two-form $B$ over large spheres surrounding the
black hole -- in our previous language, this integral is the
$Z_N$ charge divided by $N$.
The $B$ field
gives a more classical look to the hair, in that
there is a definite long-range field attached
to it.\REF\allen{T. J. Allen, M. J. Bowick, and A. Lahiri,
Phys. Lett. {\bf B237} (1990) 47.}\refmark{\bowick,\allen}
However
this appearance is deceptive: the
meaningful part of $B$
is {\it not} a classical field in the usual sense.
One sign of this,
is its observable (fractional) part cannot take arbitrarily
large values.
Indeed the fractional charge associated with
$B$ only acquires
physical significance in the interaction of the hole with
real or virtual flux tubes -- precisely the same process as
we have described in great detail above, which is
non-perturbative in $\hbar$.  Nevertheless the $B$ field is
very useful in describing the space-time process of cosmic
string lassoing a charged hole, as
we shall see.\refmark{\preskill,\cpwA}

The interpretation of ``$B$ hair'' as gauge hair depends
on the presence of the $B\wedge F$ interaction in the
Lagrangian,
which occurs in the dual theory of a massless scalar
precisely when that scalar is eaten by the vector field
according to the Higgs mechanism.  We emphasize that
$B$ hair is therefore, in the only case where we understand
it,
nothing but discrete gauge hair.  In particular it has
nothing to
do with axions, despite the unfortunate terminology
common in the literature.
One might also be tempted to consider the possibility of
global
hair, for a discrete remnant of a broken global symmetry.
Global symmetry corresponds to the absence of a $B\wedge F$
term.
We will offer
both formal and physical arguments {\it against} the
possibility
of meaningful hair in this case, however.\refmark{\cpwA}

\section{Dual charge}

Harking back to \dualra  , the charge $Q$ that couples to
the gauge field can be expressed in terms of the
antisymmetric tensor field.  The charge contained within a
region $\Omega$ is
$$
Q_\Omega=~ {e\over 2}\int_\Omega \epsilon^{ijk} \partial_i
B_{jk} d^3x
{}~=~ {e\over 2}\oint_\Sigma \epsilon^{ijk} B_{jk} d^2S_i
\equiv e\int_\Sigma B~;
\eqn\ya
$$
it is ($e$ times) the integral of the two-form $B$ over the
closed surface $\Sigma$, the boundary of $\Omega$.

Now on the black hole geometry one can have a non-zero
contribution
to the surface integral of a peculiar form.  Consider a
$B$ field that takes the form
$$
eB={Q\over 4\pi}\sin\theta d\theta\wedge d\phi~,
\eqn\closedform
$$
outside the event horizon of the hole.
This form is closed, so that the charge density $H=dB$
vanishes outside the black hole.  However if we focus on the
final entry in
\ya\ , an interesting possibility comes into view.
That is, there is the possibility, exploited precisely by
a $B$ field of the given form, of a contribution from a
bounding surface -- in our example, the event horizon --
which gets
cancelled, formally,
only by the boundary at infinity.  Thus we might expect that
the $B$ field can represent charge {\it within} the
black hole.  If so, the charge $Q$ is a type of black hole
hair.

However the physical significance of the hair thus
defined formally is not immediately obvious.  Indeed $B$
was introduced as an auxiliary quantity in the dualization
process,
and has no immediately given independent meaning.
To understand the meaning of this hair, we must consider the
role played by the term $S_{B\wedge F}$ in \Scoup\ in the
quantization of electrodynamics.  We will do so in two
stages.  First we will consider the effect of coupling $B$
to the photon in the Coulomb phase of compact
electrodynamics, the phase in which the photon is exactly
massless.  Then we will consider the effect of $B$ in a
Higgs phase, and thus arrive at a dual description of
quantum hair.

If we canonically quantize electrodynamics in the temporal
($A_0=0$) gauge, the coupling to $B$ alters the momentum
that is conjugate to the dynamical variable $A_i$; the
momentum becomes
$$
\Pi^i=E^i-{e\over 2}\epsilon^{ijk}B_{jk}~.
\eqn\conjmom
$$
This modification alters in turn the Gauss's law constraint
satisfied by the physical states, and so changes the charge
spectrum of the theory.  Let us suppose that electrodynamics
is compact, meaning that all matter fields have charges that
are integer multiples of a fundamental unit, which we will
take to be $Ne$.   Therefore, a global gauge transformation
with the gauge function $\lambda=2\pi/Ne$ must act trivially
on all states.  Now consider a gauge transformation with
$\lambda=0$ on the horizon of the black hole, and
$\lambda=2\pi/Ne$ at spatial infinity.  (The precise form of
the transformation between the horizon and infinity is
irrelevant, since all gauge transformations of compact
support act trivially on physical states.)  The action of an
infinitesimal gauge transformation on a wave-functional
$\Psi[A]$ takes the form
$$
\delta_\lambda A_i(\vec x){\delta\over \delta A_i(\vec x)}=
-\partial_j \lambda(\vec x){1\over i\hbar}
\left(E^j(\vec x)-{e\over 2}
\epsilon^{jkl}B_{kl}(\vec x)\right)~,
\eqn\gaugeaction
$$
and therefore, the condition for the finite transformation
in question to act trivially is
$$
2\pi i ~({\rm integer})=
\int d^3x\left[-\partial_j \lambda(\vec x){1\over i\hbar}
\left(E^j(\vec x)-{e\over 2}
\epsilon^{jkl}B_{kl}(\vec x)\right)\right]~.
\eqn\trivcond
$$
Now we may integrate by parts.  Since \trivcond\ must be
satisfied by any $\lambda(\vec x)$ that satisfies the stated
boundary conditions, we may discard the resulting volume
integral, and retain only the surface term
$$
2\pi i ~({\rm integer})={2\pi i\over N\hbar
e}\int_{r=\infty}
d^2 S_j\left(E^j-{e\over 2}\epsilon^{jkl}
B_{kl}\right)~.
\eqn\surfterm
$$
We then conclude that
$$
{Q\over N\hbar e}={(\rm integer)} +
{1\over N\hbar e}\left(e\int_{r=\infty}B\right)~,
\eqn\chargeshift
$$
where $Q$ has been defined as the electric flux through the
surface at infinity.

Equation \chargeshift\ clarifies the meaning of a $B$ field
that is closed but not exact.  On a space that contains
non-contractible two-spheres (like the exterior region of a
black hole) the total charge
that determines the behavior of the electric field at
infinity is not the same, in general, as the sum of all
elementary charges  that are contained in the region.  There
is an additional contribution that is classified by the
cohomology of the two form $B$.
Clearly,
we can and ought to consider the closed part of $B$ as a
means of keeping track of the
part of the charge that has ``fallen in.''

In fact, we have also found that the total charge need not
be an integer multiple of the elementary charge quantum
$N\hbar e$.
The charge associated with the $B$ field can have a
fractional part that shifts the charge spectrum away from
the integers.  This phenomenon (and the way that we have
derived it) is closely analogous to the corresponding shift
in the electric charge spectrum that is induced by a
$\theta$-term in the presence of a
magnetic monopole.\Ref\witten{E. Witten, Phys. Lett. {\bf
86B}
(1979) 283.}

\section{Dual picture of the lasso process}

Now we must understand the effect of the $B$ field in a
Higgs phase. The key is provided by \chargeshift\ -- the
surface integral of $B$ determines the fractional part of
the electric charge, the part that cannot be screened by the
Higgs condensate.  Thus dual variables provide an
alternative way to implement a $Z_N$ charge projection.
Specifying that
$$
\exp\left({2\pi i\over N\hbar}\int_{r=\infty} B\right)
= e^{i2\pi Q/N\hbar e}
\eqn\altchproj
$$
is equivalent to weighting the vorticity sectors by
appropriate phases, as described earlier.  To see this
equivalence, it is sufficient to note that the term \Scoup\
reproduces the desired phase.

It is also enlightening to consider the
Lorentzian analog of the vortex -- the spacetime process in
which a virtual string loop lassoes the black hole and then
re-annihilates.  We wish to see that \Scoup\ has the effect
of weighting this process by an appropriate Aharonov-Bohm
phase.

To be definite, suppose that the virtual string sweeps out a
sphere that encloses the black hole, and that the $B$ field
has the form \closedform.
Then we must evaluate
$$
S_{B\wedge F}/\hbar={Q\over 4\pi \hbar}
\int F_{tr}~dt dr \sin\theta d\theta d\phi~.
\eqn\fluxtubint
$$
Now we recall that, since a static cosmic string has a
magnetic field in its core, a string moving at velocity $v$
along the sphere has a radial electric field $F_{tr}$ that
is proportional to $v$.  Integrating over the cross section
of the moving string, we have
$$
\int dy dr F_{tr}=v\Phi~,
\eqn\intEfield
$$
where $\Phi$ is the magnetic flux trapped in the core of the
string, and $\hat y$ denotes the direction tangent to the
sphere and transverse to the string.  Since the string moves
in the $\hat y$ direction at velocity $v$, we may write
$dy=v dt$; thus, at each point of the sphere we obtain
$$
\int dt dr F_{tr}=\Phi~,
\eqn\tintE
$$
if the virtual string sweeps over the sphere exactly once.
Doing the $d\theta d\phi$ integral in eq.~\fluxtubint, we obtain
$$
S_{B\wedge F}/\hbar=Q\Phi/\hbar~.
$$
For a string of minimal flux, we have $\Phi=2\pi/Ne$, and so we
find
$$
e^{iS_{B\wedge F}/\hbar}=
\exp(2\pi i Q/N\hbar e)~,
\eqn\ABaction
$$
precisely the expected Aharonov-Bohm phase.

We will briefly describe another way to do the same
calculation.  First, we note that the electromagnetic field
associated with the world sheet of a flux tube of
infinitesimal thickness can be written as
$$
F_{\mu \nu} (x) ~=~ -{1\over 2}\Phi~
   \int \epsilon_{\mu \nu \rho \sigma} \epsilon^{ab}
    {\partial y^\rho \over \partial \xi^a }
      {\partial y^\sigma \over \partial \xi^b }
      \delta^4 (x - y(\xi )) d^2\xi ~,
\eqn\yg
$$
where the coordinates $(\xi^0,\xi^1)$ parametrize the world
sheet,  $y^\mu(\xi)$  is the embedding of the world sheet in
spacetime, and $\Phi$ is the magnetic flux carried by the
string in its rest frame. (Our conventions are such that
$\epsilon_{0123}=-1$.)
Now, as an example of a world sheet that wraps around the
black hole, consider a string that nucleates at the north
pole, and annihilates at the south pole, parametrized so
that
$$
\eqalign{
&\theta=\theta(\xi^0)~,\cr
&\phi=\phi(\xi^1)\cr}~.
\eqn\wsparam
$$
Then, evaluating eq.~\fluxtubint, we have
$$
S_{B\wedge F}/\hbar={Q\Phi\over 4\pi \hbar}
\int\sin\theta\partder{\theta}{\xi_0}\partder{\phi}{\xi_1}
d\xi^0 d\xi^1=Q\Phi/\hbar~.
\eqn\ABanother
$$
We find, again, the desired Aharonov-Bohm phase.

We have shown, then, that the $B\wedge F$ term in the action
has the effect of attributing appropriate phases to
processes wherein a cosmic string world sheet envelops a
charged black hole.  We explicitly demonstrated this for a
special class of field histories, but the result is general.

\section{Global symmetry and hair}

In principle one could imagine introducing dual fields
for every conserved charge in a
theory, whether or not that charge is gauged.  Formally
then, if the dual field outside a black hole is a closed
two-form that is not exact, the black hole ``carries'' the
associated charge.  However the
question arises, whether the black hole ``hair'' thus
defined
is physically observable.

Indeed the conventional wisdom is quite the contrary, namely
that
{\it continuous} global symmetries are violated (or
transcended) by black holes.
Indeed, particles that
carry globally
conserved charges can fall into a black hole.
Any information about them that is not associated with
a long-range field is completely lost when they impact the
singularity.
If the black
hole has no
internal quantum numbers, this process flouts the
conservation law, and so
the global symmetry loses its power.  In particular, if the
black hole
eventually evaporates completely and disappears, it has no
reason to
disgorge the same amount of
charge as it has swallowed.

Yet it is quite tempting to suggest that our arguments that
established the existence of quantum hair in the case where
a continuous {\it local} symmetry is broken to a nontrivial
discrete subgroup can also be applied to the case where a
continuous {\it global} symmetry symmetry is broken to a
nontrivial discrete subgroup.  In a sense, the case of a
global $U(1)$ symmetry breaking to  $Z_N$ is just a limiting
case of the Higgs mechanism that we have discussed so
extensively -- namely, the limit in which the gauge coupling
$e$ approaches zero.  And no $e$ appears in the
Aharonov-Bohm
phase $\exp(2\pi i/N)$ that is acquired by a unit
charge that circles a minimal string.  Indeed, when a global
$U(1)$ symmetry is spontaneously broken, {\it global
strings} arise, and it has recently been shown that (at
least in the $Z_2$ case) there are infinite range
interactions between particles with nontrivial discrete
charges and global strings.\Ref\jmr{J. March-Russell, J.
Preskill,
and F. Wilczek, ``Internal Frame Dragging and a Global
Analogue of the Aharonov-Bohm Effect,'' Princeton preprint
PUPT-91-1297 (1991).}

When a global $U(1)$ symmetry (or any continuous symmetry)
is spontaneously broken, the corresponding charge operator
${\tilde Q}$ is ill-defined.  But if the objects that
condense have charge $N$, then the operator $\exp(2\pi
i{\tilde Q}/\hbar N)$, which acts trivially on the
condensate, is well-defined.  (Here, ${\tilde Q}$ denotes
the classical global charge, so that a ``charge-one''
particle has ${\tilde Q}=\hbar$.)  If the global $Z_N$
charge on a black hole can really be measured via an
Aharonov-Bohm interaction with a global string, then the
same arguments that we used in Section 2 would seem to imply
that black holes can carry global $Z_N$ hair.
It would be most peculiar if this were so, for it would mean
that
the {\it breaking} of the continuous symmetry has rendered
black holes,
which previously transcended global charge conservation,
suddenly more fastidious.  Perhaps fortunately, there are
good
reasons, both formal and physical, to believe that
quantum hair associated with global symmetries does not
exist.

\pointbegin
The partition function of a black hole with specified global
$Z_N$
charge is
given by a formula like eq.~\NprojB ,
but with $Q/e$ replaced by
the global charge ${\tilde Q}$ of
the black hole, and
where $Z_k$ is
the path integral in the sector with {\it global} vorticity
$k$; \ie \ such
that the phase of the condensate rotates by $-2
\pi k$ as $\tau$ varies from $0$ to $\beta$.  Now, however,
on a fixed asymptotically flat background geometry
there are no
vortex configurations of finite action in the infinite
volume limit; in
fact, the action of a vortex diverges like $R^3$ with the
radius $R$.  Things are even worse for a vortex coupled to
gravity -- there are no asymptotically flat solutions at
all.  If we assume that such a solution exists, then we can
compute that $T^\mu_{~\mu}=T_{\tau\tau}$ is a non-zero
constant at infinity, contradicting the Einstein equations.
We conclude, then, that the $k\not= 0$ contribution to the
partition function
is snuffed out,
and with it the charge dependence.  (It may seem odd that global
vortices on the Euclidean black hole background do not exist, since
there  appears to be no reason why a virtual global string cannot wind
around the horizon of a hole.  But in fact, this winding would introduce a
physical twist in the Goldstone boson field that would asymptotically
approach the horizon at large Schwarzschild time.  The string must wind
back the other way to return the black hole to its initial state.)

\point
Alternatively, we can perform the projection onto a sector
of specified $Z_N$ charge by fixing the surface integral of
the dual two-form $B$, according to
$$
\exp\left({i\over N\hbar }\int_{r=\infty} B\right)
=e^{2\pi i{\tilde Q}/\hbar N}~.
\eqn\dualchtilde
$$
Now, when the $Z_N$ symmetry is gauged, we have seen that
the cohomology of $B$ can have a non-trivial effect that
arises from the $B\wedge F$ term in the action.  But if the
$Z_N$ symmetry is not gauged, then $B$ enters the action
only through the field strength $H=dB$ -- the closed part of
$B$ is completely decoupled.  Thus, in a different way, we
see that the partition function is independent of the
charge.

\point
The thought-experimental method of measuring the charge by
scattering global strings off a hole, and looking for an
Aharonov-Bohm type contribution to the cross section, is
less than convincing.  First of all, since there is no mass
gap in the theory, the strings are not well localized, and
even distant strings (\ie\ with distant cores)
have local interactions with the hole.  Since the
low-energy,
forward contribution to the cross-section we seek involves
slow
motion of the hole past the string, there is every
opportunity for
the black hole to disrupt or even swallow the string.
Secondly, the
Aharonov-Bohm behavior of the scattering of a string by a $Z_2$ charge
arises only if the passage of the string by the charge can be
regarded as adiabatic, as described in Ref.~\jmr.  But if we consider
a $Z_2$ charge in the vicinity of a black hole, this adiabatic condition
becomes more and more stringent as the charge gets closer and closer
to the horizon, because of the gravitational time dilation.  Thus, the
long--range interaction between charge and string is destroyed as the
charge falls into the black hole.

\point
A revealing contrast between gauged and global discrete
symmetries is found when one attempts to construct an
operator that measures discrete global charge.  In the case
of a local $Z_N$ symmetry, it is possible to construct an
operator that has support on a closed surface $\Sigma$, such
that the phase of the expectation value of this operator is
sensitive to the $Z_N$ charge enclosed
by $\Sigma$.\refmark{\AMW,\prekra}  The
existence of this operator shows that the $Z_N$ charge
cannot disappear, and so black holes must be capable of
carrying such charges.  But the attempt to construct the
corresponding operator for global $Z_N$ charge meets an
obstacle.  In the dual language, the candidate charge
operator can be expressed as
$$
{\exp\left({2\pi i\over N\hbar}\int_{\Sigma}B\right)
\over\VEV{
\exp\left({2\pi i\over N\hbar}\int_{\Sigma}B\right)}_0}~.
\eqn\propcharop
$$
Physically, we may envision this operator as an insertion of
a {\it classical} global string on the world sheet $\Sigma$,
and its expectation value may be interpreted as the
Aharonov-Bohm phase acquired by this string as it winds
around the region bounded by $\Sigma$.\refmark{\prekra}
Because the string
is poorly localized, however, the action of the string
increases with the linear size $r$ of $\Sigma$ like
$r^2\log(r)$.  Hence, it is favorable for a dynamical global
string to nucleate in order to shield the long--range effect
of the classical string; then the action is of order $r^2$
(without the logarithm).  This means that the charge deep
inside a large region is not detected.  In contrast to the
situation when the symmetry is gauged, this construction
does not force us to conclude that the charge must be
conserved when black holes are present.

\par

Finally let us add a few words concerning axions.
The first remark to be made is that the valid form
of B-hair, as
we have discussed it above, has nothing to do with axions.
The
coupling of axions to the gauge field is quite different
from
the $B\wedge F$ term.
That term, from which we derived our hair, is
inextricably linked with discrete gauge symmetries.  Whether
the
gauge symmetries are introduced in the familiar form
of $J\cdot A$ coupling, or in the less
familiar form of the $B\wedge F$ term makes no fundamental
difference,
since these are mathematically equivalent.

Even if the
$B\wedge F$ interaction is
postulated as a {\it fundamental} property of strings,
which is
the minimum necessary
to reproduce the result of the previous section, the
interpretation
of B-hair
as discrete gauge hair seems to follow ineluctably.  For
example,
in previous chapters we have seen
in detail
how effective electric fields around black holes are
built up from the virtual string looping process.
This provides a strong hint that one may work backwards, and
construct
an effective gauge theory as a {\it consequence} of the
existence of the postulated strings
and couplings.

\chapter{Primary and Secondary Hair}

The classic no-hair theorems can be interpreted in two
senses.
A {\it strong} interpretation of these theorems would say
that they
require that black holes settle down precisely to the
classic known
forms
(Kerr-Newman)
characterized completely by mass, angular momentum, and
charge.  According to this strong interpretation there can
be no
non-trivial fields outside the black hole aside from the
gravitational
and electromagnetic (or more general continuous gauge)
fields.

This strong interpretation of the no-hair theorem, however,
is
readily seen to be violated in a number of circumstances.
An example
that has been much discussed recently is the dilaton black
hole.\REF\shapere{J. Preskill, P. Schwarz, A. Shapere, S.
Trivedi,
and F. Wilczek, Mod. Phys. Lett. {\bf A6} (1991) 2353.}\REF\trivedi{A.
Shapere, S. Trivedi,
and A. Shapere, Mod. Phys. Lett. {\bf A6} (1991) 2677.}\REF\holzhey{C.
F. E. Holzhey and F. Wilczek, ``Black Holes as Elementary Particles,''
IASSNS-HEP-91/71 (1991).}\refmark{\GibMae,\Gar,\shapere -\holzhey}
Motivated by considerations in string theory, one considers
adding to
the Einstein Lagrangian additional terms of the form
$$
{\cal L_{\rm dil.}} =
\sqrt g \left({1\over 2} g^{\mu\nu}
\partial_\mu \phi \partial_\nu
\phi
-{1\over 4} e^{-2a\phi} g^{\mu\rho}g^{\nu\sigma} F_{\mu\nu}
F_{\rho\sigma} \right)\, .
\eqn\primaa
$$
This reduces to the standard Einstein-Maxwell form if $\phi
= {\rm constant}~$.
However if $g^{\mu\rho}g^{\nu\sigma} F_{\mu\nu}
F_{\rho\sigma}
\neq 0$, such as occurs outside a charged black hole, then
the
field equation for $\phi$ will not be satisfied by a
constant -- the
non-vanishing electromagnetic field acts a source for the
dilaton
field $\phi$.  Similarly an axion field, which couples
linearly to
$\epsilon^{\mu\nu\rho\sigma}F_{\mu\nu}F_{\rho\sigma}$, will
necessarily
vary, and be non-trivial, outside a
dyonic hole which carries both electric and
magnetic charge.\REF\axdyon{B. A. Campbell, N. Kaloper, and
K. A. Olive, Phys. Lett. {\bf B263} (1991) 364; B. A. Campbell,
M. J. Duncan, N. Kaloper, and
K. A. Olive, Phys. Lett. {\bf B251} (1990) 34; K. Lee and E.
J. Weinberg, ``Charged Black Holes with Scalar Hair,'' Columbia preprint
CU-TP-515 (1991).}\refmark{\trivedi,\axdyon}  Yet another example is the
recently discovered black hole embedded in a magnetic
monopole,\refmark{\Nair} which has a Higgs field core outside the
horizon.

We would like to call this sort of hair, which is generated
because
the basic fields (associated with mass, angular momentum,
and continuous
gauge charges) act as sources for other fields, {\it
secondary\/} hair.\refmark{\cpwB}

Actually, we need not invoke exotic particles to construct
examples of
secondary hair.  It occurs in the standard model, when
perturbative
corrections to the matter Lagrangian are taken into account.
For
example, perturbative corrections to the effective
Lagrangian for
the standard model can hardly fail to generate terms of the
form
$$
\Delta{\cal L} \propto \sqrt g F_{\alpha\beta}
R^\alpha_{~\rho\sigma\tau}
            R^{\beta\rho}_{~~\lambda\eta}
\epsilon^{\sigma\tau\lambda\eta} \, .
\eqn\primba
$$
Such a term will act as a source term for the
electromagnetic field in the
presence of suitable curvature.  It will, for example,
generate a
non-vanishing electric field outside a rotating
{\it electrically neutral\/} black hole:
essentially, an electric dipole moment.  (Note that for a
non-rotating
neutral black hole with no other additive quantum numbers,
the existence of
an electric field outside is forbidden by CPT.  The
correlation of
an electric dipole moment with spin direction is P and T
violating, as
of course is the interaction \primba\ responsible for it.)
Also, the magnitude of the Higgs field will not be quite
constant outside
the hole, for similar reasons.
Let $\eta$ denote the Higgs field and $v$ its expectation
value,
with $\eta \equiv v + \zeta$.
Terms in the effective Lagrangian of the
form
$$
\Delta{\cal L} \propto \sqrt g |\eta |^2
R_{\mu\nu\rho\sigma}
        R^{\mu\nu\rho\sigma} \rightarrow
        2\sqrt g v\zeta
R_{\mu\nu\rho\sigma}R^{\mu\nu\rho\sigma} + \dots
\eqn\primca
$$
then act as a linear source for $\zeta$ in the presence
of curvature, so $\zeta$ will certainly vary in a
non-trivial way
outside the hole.

Another, weaker but more profound
way of interpreting the no-hair theorems is as
statements about the {\it classification\/} of stationary
black holes.
According to this weaker interpretation, the properties of a
black
hole are completely determined, {\it within any given
theory\/}, by
the value of its mass, angular momentum, and continuous
gauge charges.
As we have seen, this weaker interpretation is violated
non-perturbatively in $\hbar$, by discrete gauge hair.  This
form
of hair expands the space of states of black holes.  It is
therefore
appropriately called {\it primary\/} hair.

In the classical analysis of the field equations, which
inspires
the no-hair theorems,
linear
perturbation theory suggests that a massless (integer) spin
$s$ field can support hair in partial waves $l\le (s-1)$.
Associated with the spin-$2$ graviton, then, we have hair in
partial waves $l=0$ (namely $M$) and $l=1$ ($J$); associated
with the spin-$1$ photon, we have hair in $l=0$ ($Q$).
We have shown that the restriction to massless fields can be
removed in the case $s=1$.  It will be interesting to see
whether
this result can be extended
to symmetries
associated with higher spins, such as are suggested by
superstring theory.  If so, then there are real prospects
that
the complete internal state of black holes (including the
wealth of states indicated  by the area law for large black
holes)
might be accounted for.

\bigskip
After this work was completed, we learned that Dowker, Gregory,
and Traschen\Ref\dowker{F. Dowker,
R. Gregory, and J. Traschen, ``Euclidean Black Hole Vortices,''
FERMILAB-Pub-91/332-A (1991).} have
also investigated the properties of the
Euclidean vortex
solutions that we discussed in Section 4.

\ack
We gratefully acknowledge helpful discussions with Mark
Alford, Martin Bucher, Gary Gibbons, Stephen Hawking,
Lawrence Krauss, John March-Russell, Alex Ridgway, Andy
Strominger, Kip Thorne, Alex Vilenkin, and Edward Witten.

\refout
\figout
\bye